\documentclass[namedreferences]{solarphysics}

\usepackage[hyperref,optionalrh,showbiblabels]{spr-sola-addons}
\usepackage{graphicx}        
\usepackage{amssymb}        
\usepackage{breakurl}        

\newcommand{\etal}{{\it et al.}}
\newcommand{\ie}{{\it i.e.}}
\newcommand{\eg}{{\it e.g.}}

\renewcommand{\vec}[1]{{\mathbfit #1}}
\newcommand{\rmd}{ {\ \mathrm d} }

\chardef\us=`\_

\begin{document}

\begin{article}
\begin{opening}
\title{Solar Plasma Radio Emission in the Presence of Imbalanced Turbulence of Kinetic-Scale Alfv\'{e}n
Waves}
\author[addressref=aff1,corref,email={olena@mao.kiev.ua}]{\inits{O.}\fnm{O.}~\lnm{Lyubchyk}\orcid{http://orcid.org/0000-0003-3993-2991}}
\author[addressref=aff2]{\inits{}\fnm{E.P.}~\lnm{Kontar}}
\author[addressref=aff3]{\inits{}\fnm{Y.M.}~\lnm{Voitenko}}
\author[addressref=aff2]{\inits{}\fnm{N.H.}~\lnm{Bian}}
\author[addressref=aff4]{\inits{}\fnm{D.B.}~\lnm{Melrose}\orcid{http://orcid.org/0000-0002-6127-4545}}
\address[id=aff1]{Main Astronomical Observatory, National Academy of Sciences of Ukraine, 27 Akademika Zabolotnoho St., 03680 Kyiv, Ukraine}
\address[id=aff2]{School of Physics and Astronomy, University of Glasgow, G12 8QQ, Scotland, UK}
\address[id=aff3]{Solar-Terrestrial Centre of Excellence, Belgian Institute for Space
Aeronomy, Ringlaan 3 Avenue Circulaire, 1180 Brussels, Belgium}
\address[id=aff4]{Sydney Institute for Astronomy, School of Physics, University of Sydney, NSW 2006, Australia}

\runningauthor{O. Lyubchyk \etal}
\runningtitle{Solar Radio Emission and Inertial Alfv\'{e}n Turbulence}

\begin{abstract}
We study the influence of kinetic-scale Alfv\'{e}nic turbulence on the
generation of plasma radio emission in the solar coronal regions where the
plasma/magnetic pressure ratio $\beta $ is smaller than the electron/ion
mass ratio $m_{e}/m_{i}$. The present study is motivated by the phenomenon
of solar type I radio storms associated with the strong magnetic
field of active regions. The measured brightness temperature of the type I storms 
can be up to $10^{10}$ K for continuum emission, and can exceed $10^{11}$ K for type
I bursts. At present, there is no generally accepted theory explaining such
high brightness temperatures and some other properties of the type I storms. 
We propose the model with the imbalanced
turbulence of kinetic-scale Alfv\'{e}n waves producing an asymmetric
quasilinear plateau on the upward half of the electron velocity
distribution. The Landau damping of resonant Langmuir waves is suppressed
and their amplitudes grow spontaneously above the thermal level. The
estimated saturation level of Langmuir waves is high enough to generate
observed type I radio emission at the fundamental plasma frequency. Harmonic
emission does not appear in our model because the backward-propagating
Langmuir waves undergo a strong Landau damping. Our model predicts $100\%$ polarization in the sense of the ordinary (o-) mode of type I emission.
\end{abstract}
\end{opening}

\section{Introduction}

\label{S-Introduction}

Numerous observations indicate the presence of Alfv\'{e}n waves throughout
the solar atmosphere (see introduction in \citealp{2015NatCo...6E7813M} and
references therein). At small scales, where the perpendicular wavelength is
close to the kinetic plasma scales (ion gyroradius $\rho _{i}$ and/or
electron inertial length $d_{e}$), Alfv\'{e}n waves transform into
dispersive Alfv\'{e}n waves (DAWs) with the following dispersion relation %
\citep{1975PhRvL..35..370H,1979JGR....84.7239G,1996JGR...101.5085L,2000SSRv...92..423S}%
:
\begin{equation}
\omega =k_{A\parallel }V_{A}\sqrt{\frac{1+k_{A\perp }^{2}\rho _{T}^{2}}{%
1+k_{A\perp }^{2}d_{e}^{2}}},  \label{1}
\end{equation}%
where $\omega $ is the wave frequency and $k_{A\parallel }$ and $k_{A\perp }$
are wave-vector components parallel and perpendicular to the mean magnetic
field $\vec{B}_{0}$; $\rho _{T}=\sqrt{1+T_{e}/T_{i}}\rho _{i}$; $\rho
_{i}=V_{Ti}/\omega _{Bi}$ is the ion gyroradius; $V_{Ti}=\sqrt{T_{i}/m_{i}}$
is the ion thermal velocity; $\omega _{Bi}=q_{i}B_{0}/\left( m_{i}c\right) $ is the
ion gyrofrequency; $d_{e}=c/\omega _{pe}=$\ $c\sqrt{m_{e}}/\sqrt{4\pi
n_{0}\left\vert q_{e}\right\vert }$ is the electron inertial length; $V_{A}=B_{0}/%
\sqrt{4\pi n_{0}m_{i}}$ is the Alfv\'{e}n velocity; $n_{0}$ is the plasma number
density; and $q_{s}$, $m_{s}$, and $T_{s}$ are the elementary charge, mass, and
temperature of particles ($s=i$ for the ions and $s=e$ for the electrons). 
Note that in equations $T_{s}$ is expressed in energy units.
We consider a hydrogen plasma where $q_{i}=\left\vert q_{e}\right\vert =e$,
with $e$ is the proton charge. DAWs are called inertial Alfv\'{e}n waves
(IAWs) when the inertial term dominates in Equation (\ref{1}), and 
kinetic Alfv\'{e}n waves (KAWs) when the kinetic term dominates in Equation (%
\ref{1}).

The inertia of electrons plays a significant role in rarefied plasmas where
the magnetic field is strong enough to make $\beta <m_{e}/m_{i}\ll 1$. On
the other hand, in an intermediate beta plasma ($m_{e}/m_{i}\ll \beta \ll 1)$
the effects due to a finite Larmor radius become dominant in DAWs. Recent
observations have clearly demonstrated that the DAW turbulence exists in
both the $\beta >m_{e}/m_{i}$ plasma environments, such as the solar wind %
\citep{2012ApJ...745L...8H}, and in the $\beta <m_{e}/m_{i}$ plasma
environments, such as the auroral zones below 4 Earth radii %
\citep{2008PhRvL.100q5003C}. DAWs cannot be observed directly in the solar
corona. However, spectral line observations indicate that there are plenty
of unresolved nonthermal motions implying Alfv\'{e}n waves with velocity
amplitudes $\approx 30 $ km s$^{-1}$ and higher \citep{1998A&A...339..208B}.

The parallel electric field of DAWs makes them efficient in Cherenkov
resonant interaction with plasma particles. The ion and electron dynamics in
the solar wind under the influence of KAW turbulence was examinated by %
\citet{2012PhPl...19d2704R}. It was shown that the particle diffusion
governed by KAWs leads to the development of a plateau in the electron
distribution and a step-like distribution for the superthermal ions. These
distributions are found to be unstable to electromagnetic waves due to ion
cyclotron resonance. The analytic theory of proton diffusion driven by the
KAW spectra observed in the solar wind was developed by %
\citet{2013SoPh..288..369V}. These authors performed kinetic simulations of the
velocity-space diffusion of protons \citep{2013SoPh..288..355P}. It was
shown that the presence of Alfv\'{e}nic turbulence in the solar wind leads
to a fast development of nonthermal tails in the proton velocity
distribution function. The tails are already noticeable at distances about
one solar radius from the simulation boundary, and they increase rapidly
with radial distance and become pronounced beyond 2--3 solar radii.

Note that DAWs can resonantly exchange energy with particles in
the velocity range between the Alfv\'{e}n velocity $V_{A}$ and the
electron thermal velocity $V_{Te}$. As in the solar atmosphere
usually $V_{A}<V_{Te}$, and DAWs are usually not efficient agents for resonant
acceleration of coronal electrons to suprathermal energies 
(see however \citealp{2016A&A...589A.101A}). The favourable 
conditions for the suprathermal electron acceleration are in the regions with
strong magnetic fields and/or low temperatures, where the electron thermal
speed drops under the Alfv\'{e}n speed, $V_{A}>V_{Te}$. In such
regions, DAW turbulence leads to the diffusive acceleration of resonant
suprathermal electrons to Alfv\'{e}nic velocities creating a quasi-linear
plateau on the electron velocity distribution.

The purpose of this paper is to study possible effects of the IAW
turbulence in the generation of solar coronal radio emission. The proposed
scenario is as follows. First, the IAW turbulence produces a flat
(plateau-like) velocity distribution of the electrons in the velocity range 
$\sqrt{1+T_{i}/T_{e}}V_{Te}<V_{\parallel }<V_{A}$. Such local
flattening of the velocity distribution suppresses the Landau damping of the
resonant Langmuir waves (LWs), making possible their spontaneous growth to
high nonthermal levels. Then the resulting high-amplitude LWs can interact
nonlinearly with low-frequency plasma waves generating electromagnetic
radiation (radio waves) close to the local plasma frequency. 

Our scenario can contribute to radio emission from the coronal
regions where the plasma/magnetic pressure ratio $\beta $ is smaller than
the electron/ion mass ratio $m_{e}/m_{i}$. In particular, this mechanism can
account for type I solar radio storms -- the most common manifestations
of solar radio emission at meter wavelengths %
\citep{1977sns..book.....E,1985srph.book.....M}.

The outline of the paper is as follows. The influence of the IAW turbulence on
the electron velocity distribution and spectral energy density of Langmuir
waves are estimated in Section~\ref{S-W_L}. The theory for fundamental
plasma emission by the fusion/decay processes $L\pm S\rightarrow T$ is
presented in Section~\ref{S-T_T}, including the rate equations for the
emission processes (Section~\ref{subS-rate equations}), the saturation level of
fundamental plasma emission (Section~\ref{subS-sat_level}) and its
absorption during propagation (Section~\ref{subS-absorption}). Possible
application to type I solar radio bursts is discussed in Section~\ref%
{S-Application}. Our main conclusions are given in Section~\ref%
{S-Conclusions}.

\section{Spectral Energy Density of LWs due to IAW Turbulence}

\label{S-W_L}

The Landau resonance with electrons is made possible by the existence of a
parallel electric field which is the result of the combined effect of the
electron pressure gradient and electron inertia in Ohm's law. This parallel
electric field is also the very reason for the wave dispersion (see Equation
(\ref{1})) in a collisionless plasma. In the case $\rho_ T= d_e$, the wave
becomes dispersionless and the wave-particle resonance reduces to a single
point in velocity-space (see Appendix A for the details). Otherwise the
range of Landau resonance in velocity-space is finite. The generation of a
spectrum of parallel electric field fluctuations by Alfv\'{e}nic turbulence
was studied in \citet{2010A&A...519A.114B}. Parallel electric field
amplification and spectral formation by phase mixing of Alfv\'{e}n waves
were studied in \citet{2011A&A...527A.130B} and the role of this electric
field in the bulk energization of electrons during solar flares was developed
in \citet{2014SoPh..289..881M}.

The Landau resonance between electrons and DAWs occurs when the electron
velocity $V_{\parallel }$ is equal to the wave phase speed $V_{\mathrm{DAW}}=\omega
/k_{A\parallel }$, \ie
\[
V_{\parallel }=V_{A}\sqrt{\frac{1+k_{A\perp }^{2}\rho _{T}^{2}}{1+k_{A\perp
}^{2}d_{e}^{2}}}.
\]%
The range of resonant velocities extends from $V_{\parallel }=$ $V_{T}=\sqrt{%
1+T_{i}/T_{e}}V_{Te}$ at $k_{A\perp }\rightarrow \infty $ to $V_{\parallel
}=V_{A}$ at $k_{A\perp }\rightarrow 0$. Note that we are interested in the
regions where the plasma $\beta $ can be smaller than the electron to ion
mass ratio $m_{e}/m_{i}$. In such a low-$\beta $ plasma the DAWs are IAWs, and
phase velocities are larger than the electron thermal velocity $V_{Te}$.

\begin{figure}[tbp]
\includegraphics[width=7.5cm]{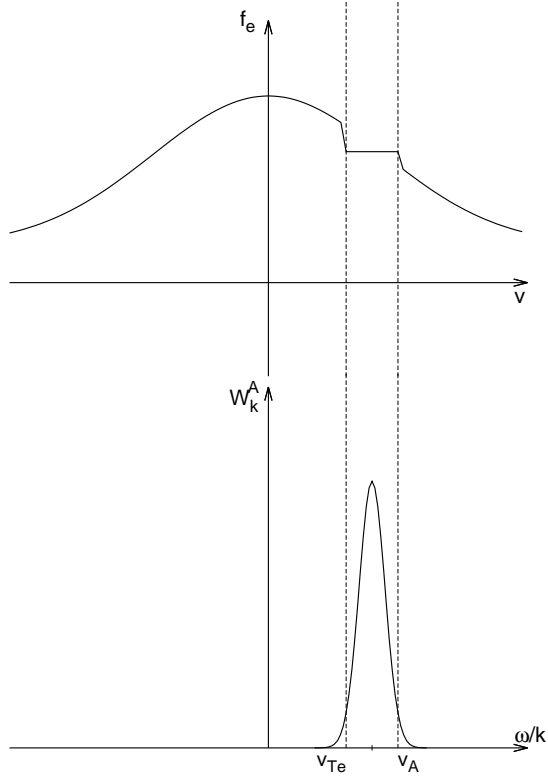}
\caption{Illustration of the time-asymptotic electron velocity
distribution $f_{e}$ modified by the IAW spectrum $W_{k}^{A}$. The quasilinear diffusion establishes a plateau on the initially
Maxwellian distribution in the resonant velocity range $\protect\sqrt{%
1+T_{i}/T_{e}}V_{Te}<V_{\parallel }<V_{A}$. Outside this interval
the velocity distribution remains Maxwellian.}
\label{fig:schem}
\end{figure}

Starting from the initially Maxwellian velocity distribution, the
quasi-linear plateau can be formed by IAWs in the resonant velocity range $
\sqrt{1+T_{i}/T_{e}}V_{Te}<\left\vert V_{\parallel }=V_{\mathrm{IAW}}\right\vert
<V_{A}$, as is sketched in Figure \ref{fig:schem}. This process is
essentially the same as the quasilinear evolution of initially Maxwellian
velocity distributions driven by KAWs in the solar wind, which has been studied
by \citet{2012PhPl...19d2704R}, \citet{2013SoPh..288..369V}, \citet{2013SoPh..288..355P}. The particle beams are not needed for that. In turn, the
reduced velocity-space gradient within the plateau suppresses the linear
Landau damping of resonant LWs with phase velocities $\sqrt{1+T_{i}/T_{e}}%
V_{Te}<$ $V_{LW}<V_{A}$. As a consequence, a spontaneous
growth of LW amplitudes is possible in this velocity range.

We have to note that the initial velocity distribution of the electrons does not have to be exactly Maxwellian, it can be {\eg} a kappa-distribution often observed in space. However, this would not affect the IAW dispersion and polarization, which depend on the bulk plasma parameters, and the basic physical picture would be the same. First, the IAWs flatten the electron velocity distribution in the resonant velocity range, which is then followed by the spontaneous growth of the Langmuir wave amplitudes above the thermal level. 

In-situ satellite observations have revealed that the MHD Alfv\'{e}nic
turbulence in the solar wind is dominated by the anti-sunward wave flux (%
\citealp{2013LRSP...10....2B}, and references therein). The corresponding
wave turbulence is referred to as imbalanced. \citet{1999ApJ...523L..93M}
suggested that the imbalanced turbulence of MHD Alfv\'{e}n waves develops
also in the solar corona, where it can reach dissipative scales and heat
plasma. \citet{2016ApJ...832L..20V} investigated the MHD-kinetic turbulence
transition and showed that the KAW turbulence is less imbalanced than the
parent MHD turbulence but remains imbalanced throughout. Other possible
sources for the imbalanced KAW turbulence in the solar corona include phase
mixing \citep{2000A&A...357.1073V} and magnetic reconnection %
\citep{1998SoPh..182..411V}. In solar flares these KAWs can lead to 
impulsive plasma heating \citep{1998SoPh..182..411V} and non-local electron
acceleration to relativistic energies \citep{2016A&A...589A.101A}.

Contrary to the balanced IAW turbulence, which is symmetric with
respect to $k_{\parallel }\rightarrow -k_{\parallel }$, the wave
amplitudes in the imbalanced IAW turbulence are much smaller at $%
k_{\parallel }<0$ than at $k_{\parallel }>0$. The balanced
IAW turbulence forms two symmetric quasilinear plateaus in the electron velocity distribution function (VDF),
at $-V_{A}<$ $V_{\parallel }<$ $-\sqrt{1+T_{i}/T_{e}}
V_{Te}$ and $\sqrt{1+T_{i}/T_{e}}V_{Te}<$ $V_{\parallel
}<$ $V_{A}$. The imbalanced IAW turbulence forms the
plateaus asymmetrically, preferentially in the propagation direction of the
dominant IAW component, which we take as positive, {\ie} at $\sqrt{%
1+T_{i}/T_{e}}V_{Te}<$ $V_{\parallel }<$ $V_{A}$. The spontaneous emission of LWs can be studied independently for forward
and backward plateaus. 

As the origin of coronal Alfv\'{e}n waves is at the coronal base, it
is natural to expect that the upward wave flux in the solar corona is larger
than the downward one and the corresponding turbulence is imbalanced. The
turbulence imbalance has a striking consequence in our scenario. As the
quasilinear plateau is formed by the IAW turbulence preferentially in the
direction of the dominant upward-propagating IAW fraction, {\ie} at $\sqrt{%
1+T_{i}/T_{e}}V_{Te}<V_{\parallel }<V_{A}$, the Landau damping of
upward-propagating Langmuir waves is highly reduced in this velocity range
and they can grow spontaneously to high amplitudes. On the contrary, the
downward IAW flux is weaker and not so efficient in forming the plateau, in
which case the Landau damping of downward Langmuir waves remains strong and
prevents their spontaneous growth. In these conditions, there are no
counter-propagating Langmuir waves and only fundamental radio emission can
be generated by the plasma emission mechanism.

In the 1D approximation, the governing equation for the spectral energy
density of LWs is %
\citep{1963JETP...16..682V,1969JETP...30..131R,1987ApJ...321..721H}
\begin{equation}
\frac{\rmd W_{k}^{l}}{\rmd t}=\frac{\pi \omega _{pe}^{3}}{n_ek^{2}}W_{k}\left[ \frac{%
\partial f}{\partial V}\right] _{V=\omega _{pe}/k}+\frac{\omega
_{pe}^{3}m_{e}}{4\pi n_{e}}\left[ V\ln \left( \frac{V}{V_{Te}}\right) f%
\right] _{V=\omega _{pe}/k}-\gamma _{\mathrm{coll}}W_{k}^{l},  \label{W_in}
\end{equation}%
where $W^{l}\left( k,t\right) $ [ergs cm$^{-2}$] is the spectral energy
density of LWs, $f\left( V,t\right) $ [electrons cm$^{-3}$ (cm/s)$^{-1}$] is
the electron distribution function, $k$ is the LW wavenumber (we are mostly
interested in parallel-propagating LWs with $k\equiv k_{\parallel }$), $%
V $ is the parallel component of the particle velocity $\left( V\equiv
V_{\parallel }\right) $, and $\omega _{pe}$ is the local electron plasma
frequency. Here the following normalisations are used:
\[
\int\limits_{-\infty }^{\infty }f\rmd V=n_e,
\]%
where $n_{e}$ is the density of the background plasma, in cm$^{-3}$ and
\[
\int\limits_{-k_{De}}^{k_{De}}W_{k}^{l}\rmd k=W^{l},
\]%
where $W^{l}$ is the total energy density of the waves in ergs cm$^{-3}$, with $%
k_{De}=\omega _{pe}/V_{Te}.$ The first term on the right-hand side (RHS) of
Equation (\ref{W_in}) describes the induced absorption of plasma waves by
electrons (Landau damping) and the second term describes the spontaneous
emission. This emission (and the corresponding absorption) is resonant,
meaning an electron at velocity $V$ interacts only with a wave at wavenumber
$k = \omega_{pe}/V$. Collisional damping of LWs is given by the third term
on the RHS, where $\gamma _{\mathrm{coll}}=\frac{1}{3}\sqrt{\frac{2}{\pi }}\frac{%
\Gamma }{V_{Te}^{3}}\simeq \Gamma /\left( 4V_{Te}^{3}\right) $ with $\Gamma
=4\pi e^{4}n_{e}\ln \Lambda /m_{e}^{2}$ where $\ln \Lambda $ is the Coulomb
logarithm (approximately 20 in the solar corona).

According to Equation (\ref{W_in}), in the resonant range the spectral energy
density of LWs is
\begin{equation}
W_{k}^{l}=\frac{1}{\gamma _{\mathrm{coll}}}\frac{\omega _{pe}^{3}m_{e}}{4\pi n_{e}}%
\left[ V\ln \left( \frac{V}{V_{Te}}\right) f\right] _{V=\omega _{pe}/k}.
\end{equation}%
The local plateau function $f_{\mathrm{\mathrm{pl}}}$ can be found from conservation of
resonant particles: $\int\limits_{V_{\min }}^{V_{\max }}\left[ f\left(
V,\infty \right) -f\left( V,0\right) \right]\rmd V=0,$ and $f\left( V,\infty
\right) =\mathrm{const}$. Following \citet{2013SoPh..288..369V}, we express $f_{\mathrm{pl}}$
in terms of an error function:
\begin{eqnarray}
f_{\mathrm{pl}} &=&\frac{1}{\Delta V}\int\limits_{V_{\min }}^{V_{\max }}f_{M}\rmd V=
\label{f_pl} \\
&=&\frac{n_{e}}{\Delta V}\frac{\mathrm{erf}\left( V_{\max }/\sqrt{2}%
V_{Te}\right) -\mathrm{erf}\left( V_{\min }/\sqrt{2}V_{Te}\right) }{2},
\nonumber
\end{eqnarray}%
where $\Delta V=V_{\max }-V_{\min }$, $V_{\max }\left( V_{\min }\right) $ is
the maximum (minimum) velocity of the electrons involved in the plateau, and
$f_{M}$ is the Maxwellian distribution. Since in the solar corona $\beta $
cannot be much smaller than $m_{e}/m_{i}$, the IAWs resonate within a narrow
range of phase velocities and quasilinear diffusion should be fast enough to
create the plateau from $V_{\min }\approx $ $\sqrt{1+T_{i}/T_{e}}V_{Te}$ to $%
V_{\max }\approx V_{A}$.

Consequently, the normalized spectral energy of LWs takes the following
form:
\begin{equation}
\frac{W_{k}^{l}}{W_{\mathrm{th}}^{l}}=\pi \frac{V_{Te}}{\Delta V}\frac{\omega _{pe}}{%
\gamma _{\mathrm{coll}}}\frac{1}{k^{3}\lambda _{De}^{3}}\frac{\mathrm{erf}\left(
V_{A}/\sqrt{2}V_{Te}\right) -\mathrm{erf}\left( \sqrt{1+T_{i}/T_{e}}/\sqrt{2}%
\right) }{2},  \label{W_Lplatnorm}
\end{equation}%
where the thermal level of LWs is given by
\citep{2012A&A...539A..43K,
2014A&A...562A..57R}:
\begin{eqnarray}
W_{\mathrm{th}}^{l} &=&\frac{T_{e}}{4\pi ^{2}}\frac{k^{2}\ln \left( \frac{1}{k\lambda
_{De}}\right) }{1+\frac{\gamma _{coll}}{\omega _{pe}}\sqrt{\frac{2}{\pi }}%
k^{3}\lambda _{De}^{3}\exp \left[ \frac{1}{2k^{2}\lambda _{De}^{2}}\right] }
\label{W_fultherm} \\
&\simeq &\frac{T_{e}}{4\pi ^{2}}k^{2}\ln \left( \frac{1}{k\lambda _{De}}%
\right) .  \nonumber
\end{eqnarray}

\begin{figure*}[tbp]
\includegraphics[angle=0, width=10cm]{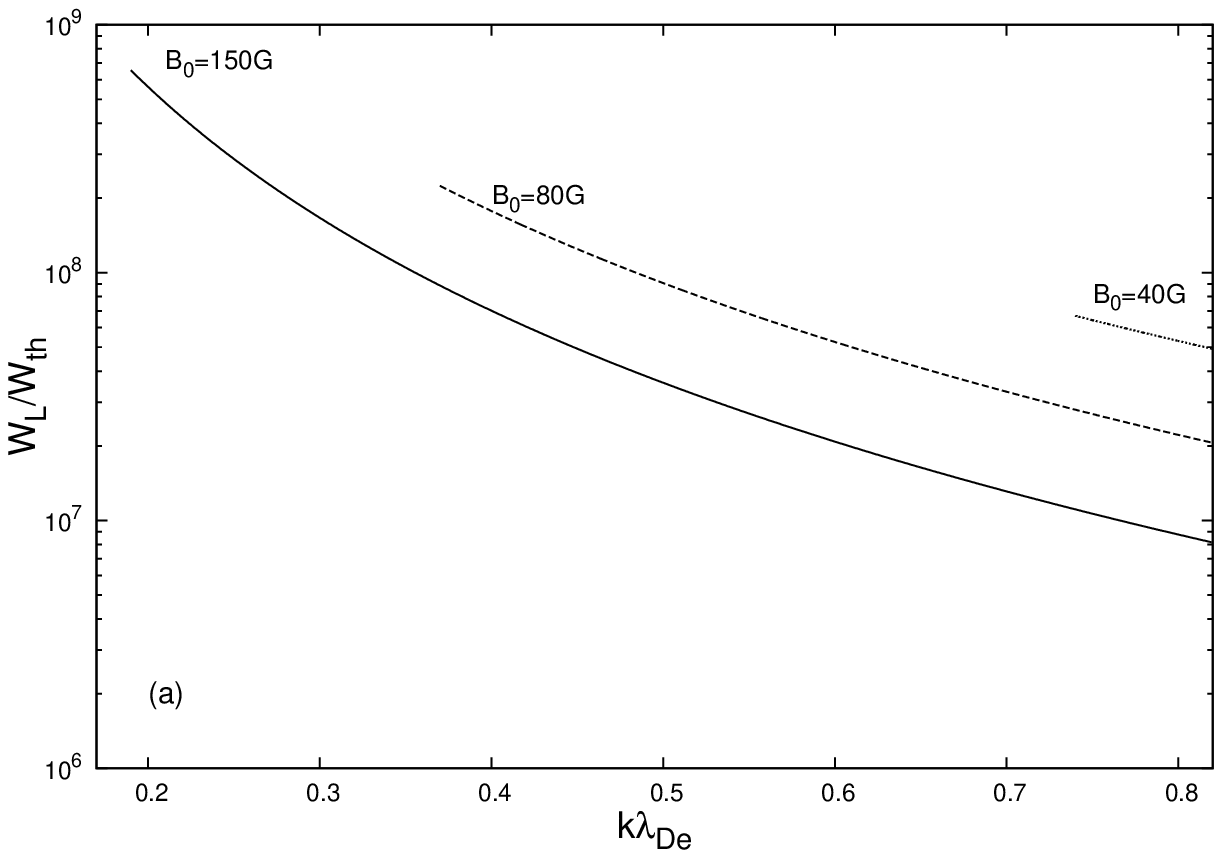} \includegraphics[angle=0,
width=10cm]{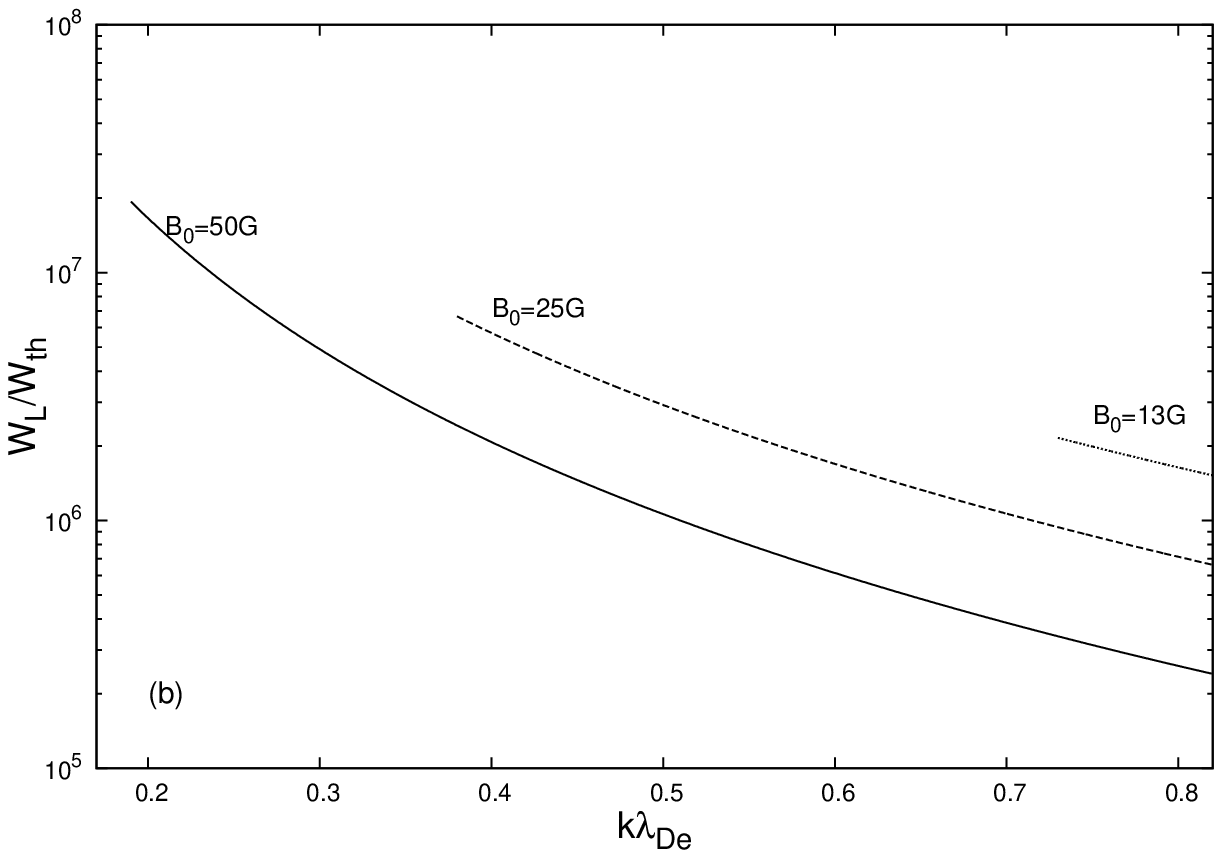}
\caption{Normalized spectral energy density of Langmuir waves in the
resonant wavenumber range $V_{Te}/V_{A}<k\protect\lambda _{De}<0.82$. The
plasma parameters are $n_{0}=2.8\times 10^{8}$ cm$^{-3}$; $T_{i}/T_{e}=0.5$.
Case (a) corresponds to $T_{e}= 10^{6}$ K and three values of magnetic field
strength $B_0 = 150$, $80$ and $40$ G. Case (b) corresponds to $T_{e}= 10^{5}
$ K and three values of magnetic field strength $B_0 = 50$, $25$ and $13$ G.}
\label{fig:W_Lp}
\end{figure*}

Figure \ref{fig:W_Lp} shows the normalized spectral energy density
of Langmuir waves in the resonant wavenumber range $V_{Te}/V_{A}<k\lambda
_{De}<0.82$. We have considered two types of possible coronal
plasma parameters for the heliocentric distances $R/{\mathrm R_{\odot }}\approx  1.1$
with $n_{0}=2.8\times 10^{8}$ cm$^{-3}$ or $f_{pe}=150$ MHz %
\citep{2005A&A...435.1123W}. The first one has $T_{e}=10^{6}$ K and
a stronger magnetic field in the range $B_{0}=150$--$40$ G (Figure \ref%
{fig:W_Lp}a). The second one has a lower temperature $T_{e}=10^{5}$ K and
a weaker magnetic field in the range $B_{0}=50$--$13$ G (Figure \ref{fig:W_Lp}b). Such values for $B_{0}$ from 10 to 150 G and $T_{e}$ from $10^{5}$ K
to $10^{6}$ K (the temperature could be even lower, down to $10^{4}$ K) are
supported by recent observations (see papers by \citealp{2016ApJ...833....5S}; \citealp{2015ApJ...806...81A}) for the relevant heliocentric distances $R/{\mathrm R_{\odot }}\approx  1.1$.
As is seen from Figure \ref{fig:W_Lp}, the normalized spectral energy
density of LWs reaches its maximum value
\begin{equation}
W_{k}^{l}/W_{\mathrm{th}}^{l}\simeq 6\times 10^{8},  \label{W_k/W_th_6}
\end{equation}%
for $B_{0}=150$ G, $T_{e}=10^{6}$ K and
\begin{equation}
W_{k}^{l}/W_{\mathrm{th}}^{l}\simeq 2\times 10^{7},  \label{W_k/W_th_5}
\end{equation}%
for $B_{0}=50$ G, $T_{e}=10^{5}$ K and for the resonant velocities of LWs close
to the upper boundary, $V_{LW}\lesssim V_{A}$, which correspond to LW
wavenumbers $k\lambda _{De}\gtrsim 0.19$. For lower values of magnetic field
strength the normalized spectral energy density of LWs is still high,
whereas the resonance wavenumber range becomes narrower.

Such a high level of LWs is obtained under the conditions that the
quasilinear Landau damping of LWs (first term in the RHS of Equation (\ref%
{W_in})) is weaker than their collisional damping (third term). When the
slope of the plateau is large enough to make the Landau damping dominant,
the spontaneous growth of LWs is reduced. Using results by %
\citet{2006SSRv..122..255V}, we estimated that this first term is
smaller than the third one for realistic IAW parameters and, consequently,
the state of saturation is governed by the balance between the second and
third terms.

\section{Fundamental Plasma Emission from the IAW Turbulent Plasma}

\label{S-T_T}

\subsection{Kinetic Equations for the Spectral Energy Density of Transverse
Waves due to the Process $L\pm S\rightarrow T$}

\label{subS-rate equations}

We consider the theory of fundamental plasma emission generated by the
nonlinear fusion $L+S\rightarrow T$ (hereafter f- process), and/or
decay $L\rightarrow T+S$ (hereafter d- process). Here $L$ is the
Langmuir wave, $S$ is the ion sound wave  (ISW) and $T$ is the transverse
radio wave. The kinetic equations for fundamental emission by the processes $%
L\pm S\rightarrow T$ are based on the general expressions in
\citet{1980gbs..bookR....M} and \citet{1995lnlp.book.....T}
\begin{eqnarray}
&&\frac{\rmd W_{T\pm }\left( \vec{k}_{T}\right) }{\rmd t}=\frac{\omega _{T}}{%
\hbar }\int \frac{\rmd \vec{k}_{L}\rmd \vec{k}_{S}}{\left( 2\pi
\right) ^{3}}u_{TLS\pm }\left( \vec{k}_{T},\vec{k}_{L},\vec{k}%
_{S}\right) \times   \label{kin_eq} \\
&&\left[ \frac{W_{L}}{\omega _{L}}\left( \frac{W_{S}}{\omega _{S}}\mp \frac{%
W_{T}}{\omega _{T}}\right) -\frac{W_{T}}{\omega _{T}}\frac{W_{S}}{\omega _{S}%
}\right] ,  \nonumber
\end{eqnarray}%
with the following emission probability
\[
u_{TLS\pm }=C\frac{\left[ \vec{k}_{T}\times \vec{k}_{L}\right]
^{2}}{k_{T}^{2}k_{L}^{2}}\delta \left( \vec{k}_{T}-\vec{k}_{L}\mp
\vec{k}_{S}\right) \delta \left( \omega _{T}-\omega _{L}\mp \omega
_{S}\right) ,
\]%
where
\[
C=\frac{\hbar e^{2}\left( 2\pi \right) ^{6}}{8\pi m_{e}^{2}V_{Te}^{4}}\frac{%
\omega _{pe}^{3}\omega _{S}^{3}}{\omega _{pi}^{2}k_{s}^{2}\omega _{T}},
\]%
and the ``$\pm $'' signs refer to the f- and d- processes,
respectively. The delta functions lead to kinematic constraints on the
frequencies and wave vectors of waves taking part in the considered
processes. According to the energy and momentum conservation conditions
\[
\omega _{L}\left( \vec{k}_{L}\right) \pm \omega _{S}\left( \vec{k}%
_{S}\right) =\omega _{T}\left( \vec{k}_{T}\right) ,
\]%
\[
\vec{k}_{L}\pm \vec{k}_{S}=\vec{k}_{T},
\]%
the wavenumbers of interacting waves must satisfy $k_{S}\simeq \mp k_{L}$
and the electromagnetic emission is generated approximately perpendicular to
the initial LW with the wavenumber
\begin{equation}
k_{T}d_{e}\approx \sqrt{3}k_{L}\lambda _{De}.
\end{equation}%
Using an angle averaged emission model
\citep{2014A&A...562A..57R,
2013PhDT.......293R}, given in detail in Appendix B, we get the following
equation for the spectral energy density of transverse waves:
\begin{eqnarray}
&&\frac{\rmd W_{T\pm }^{Av}\left( k_{T}\right) }{\rmd t}=\frac{\pi \omega
_{pe}^{4}V_{s}}{24n_{e}T_{e}V_{Te}^{2}}\left( 1+\frac{3T_{i}}{T_{e}}\right)
\sqrt{1+k_{T}^{2}d_{e}^{2}}\times  \\
&&\left[ \frac{W_{L}^{Av}\left( k_{L}\right) }{\omega _{L}}\frac{4\pi
k_{T}^{2}}{\Delta \Omega k_{S}^{2}}\frac{W_{S}^{Av}\left( k_{S}\right) }{%
\omega _{S}}\mp \frac{W_{L}^{Av}\left( k_{L}\right) }{\omega _{L}}\frac{%
W_{T}^{Av}\left( k_{T}\right) }{\omega _{T}}-\frac{W_{T}^{Av}\left(
k_{T}\right) }{\omega _{T}}\frac{W_{S}^{Av}\left( k_{S}\right) }{\omega _{S}}%
\right].  \nonumber
\end{eqnarray}%
Rewriting this in terms of the brightness temperature
\begin{equation}
k_{B}T_{k}^{T}=\frac{2\pi ^{2}W_{Tk}}{k_{T}^{2}},
\end{equation}%
we obtain the following basic equation for the transverse wave brightness
temperature induced by $L\pm S\rightarrow T$ processes:
\begin{eqnarray}
&&\frac{\rmd T_{T\pm }\left( k_{T}\right) }{\rmd t}=\frac{\pi \omega _{pe}^{4}V_{s}}{%
24n_{e}T_{e}V_{Te}^{2}}\left( 1+\frac{3T_{i}}{T_{e}}\right) \sqrt{%
1+k_{T}^{2}d_{e}^{2}}\times   \label{T_k} \\
&&\left[ \frac{\eta }{k_{S}^{2}}\frac{W_{L}^{Av}\left( k_{L}\right) }{\omega
_{L}}\frac{W_{S}^{Av}\left( k_{S}\right) }{\omega _{S}}\mp \frac{%
W_{L}^{Av}\left( k_{L}\right) }{\omega _{L}}\frac{T_{T}\left( k_{T}\right) }{%
\omega _{T}}-\frac{T_{T}\left( k_{T}\right) }{\omega _{T}}\frac{%
W_{S}^{Av}\left( k_{S}\right) }{\omega _{S}}\right] ,  \nonumber
\end{eqnarray}%
where
\[
\eta =\frac{\left( 2\pi \right) ^{3}}{\Delta \Omega },
\]%
and
\[
k_{S}^{2}=k_{T}^{2}+k_{L}^{2};k_{L}\approx \mp \frac{k_{T}d_{e}}{\sqrt{3}%
\lambda _{De}}.
\]

\subsection{Saturation Level of Plasma Emission in Presence of Nonthermal
Level of Langmuir and Ion-Sound Waves}

\label{subS-sat_level}

The condition $\rmd T_{T\pm }\left( k_{T}\right) /\rmd t=0$ in the expression (\ref%
{T_k}) defines the saturation of the f- and d- processes.
Therefore, the saturation brightness is
\begin{equation}
T_{T_{+}}=\omega _{T}\frac{\frac{\eta }{k_{S}^{2}}\frac{W_{L}}{\omega _{L}}%
\frac{W_{S}}{\omega _{S}}}{\frac{W_{S}}{\omega _{S}}+\frac{W_{L}}{\omega _{L}%
}},  \label{sat_f}
\end{equation}%
for the f- process and
\begin{equation}
T_{T_{-}}=\omega _{T}\frac{\frac{\eta }{k_{S}^{2}}\frac{W_{L}}{\omega _{L}}%
\frac{W_{S}}{\omega _{S}}}{\frac{W_{S}}{\omega _{S}}-\frac{W_{L}}{\omega _{L}%
}},  \label{sat_d}
\end{equation}
for the d- process.

Let us now discuss separately two cases i) $W_{S}/\omega _{S}\ll
W_{L}/\omega _{L}$ and ii) $W_{S}/\omega _{S}\gg W_{L}/\omega _{L}$.

\subsubsection{Case i): $W_{S}/\protect\omega _{S}\ll W_{L}/\protect\omega %
_{L}$}

First we consider the saturation level of the processes for the non-thermal
level of LWs due to the spontaneous emission Equation (\ref{W_Lplatnorm})
and thermal level of ion-sound waves, which is defined by:
\begin{equation}
W_{S\mathrm{th}}=T_{e}k_{De}^{2}\frac{k_{De}^{2}}{k_{De}^{2}+k^{2}}.
\end{equation}

\begin{figure*}[h!]
\includegraphics[angle=0, width=10cm]{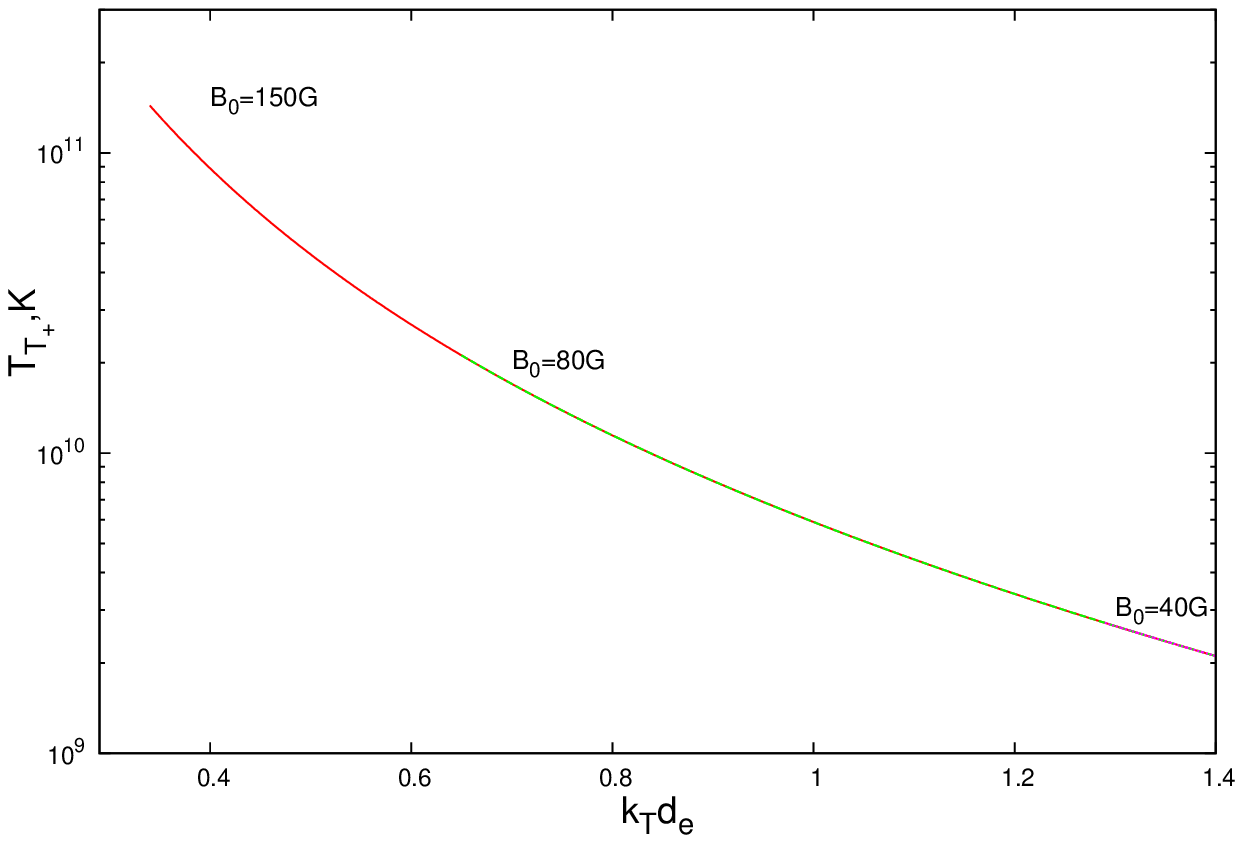} 
\includegraphics[angle=0, width=10cm]{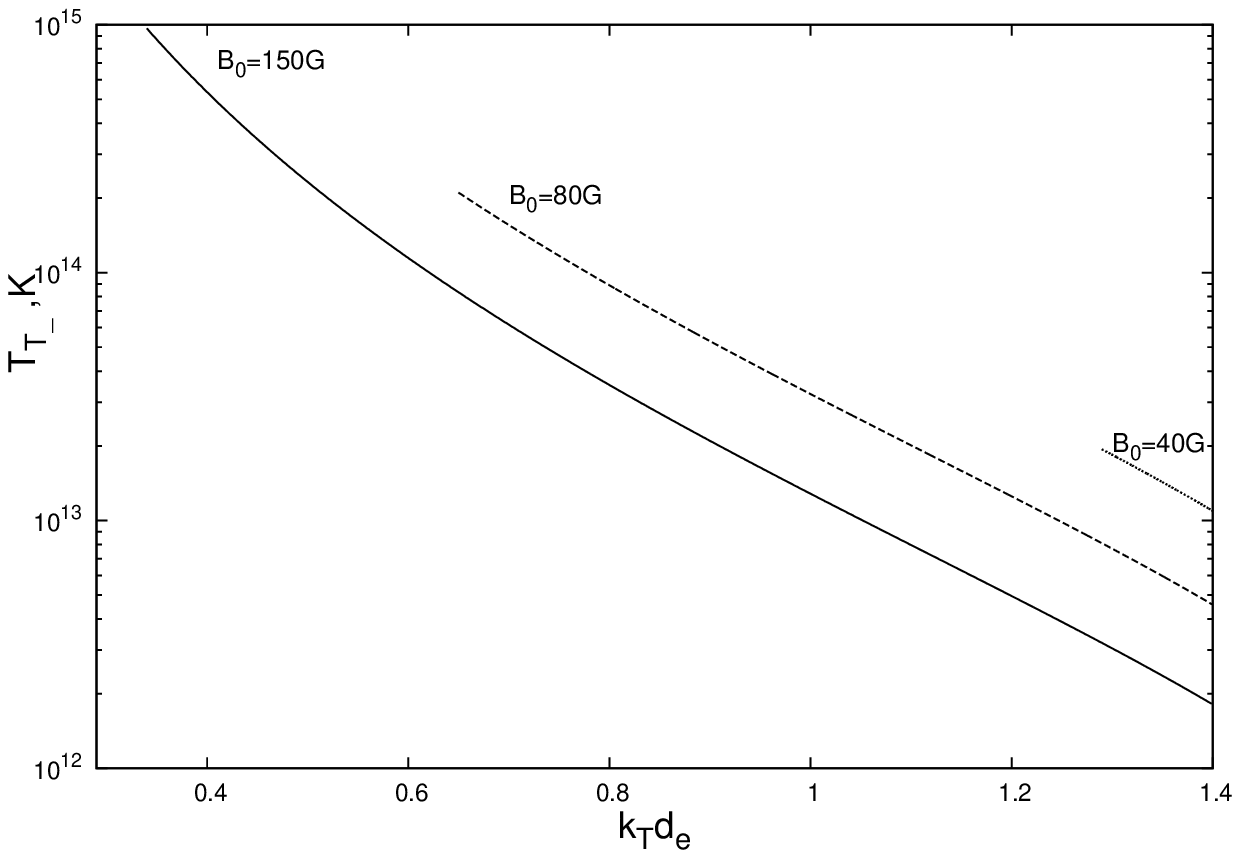} 
\caption{ Brightness temperature $T_T(k)$ of fundamental radio emission for the
f- process (top panel), and d- process (bottom panel). Here we
consider a nonthermal level of LWs, thermal ISWs and $\protect\eta=4\protect%
\pi ^{2}$. The plasma parameters are $n_{0}=2.8\times 10^{8}$ cm$^{-3}$; $%
T_{e}=10^{6}$ K; $T_{i}/T_{e}=0.5$ and three values of magnetic field
strength $B_0 = 150$, $80$ and $40$ G.}
\label{fig:satfdLpT_6}
\end{figure*}

As in this case $W_{L}\gg W_{S}$, then accordingly to Equation (\ref{sat_f})
$T_{T+}$ saturates at about $W_{Sth}$ for the f- process$:$
\begin{equation}
T_{T+}=\omega _{T}\frac{\eta }{k_{S}^{2}}\frac{W_{Sth}}{\omega _{S}}.
\end{equation}

The saturation level $T_{T-}$ for the d- process $L\rightarrow T+S$
is defined by Equation (\ref{sat_d}). Consequently, for $W_{L}\gg W_{S},$ we
should get exponential growth causing both $T_{T}$ and $W_{S}$ to
increase until $W_{S}\gg W_{L}$ when the process saturates at the level
\begin{equation}
T_{T-}=\omega _{T}\frac{\eta }{k_{S}^{2}}\frac{W_{L}}{\omega _{L}},
\label{dsatL}
\end{equation}
with $W_{L}$ defined by Equation (\ref{W_Lplatnorm}).

\begin{figure*}[h!]
\includegraphics[angle=0, width=10cm]{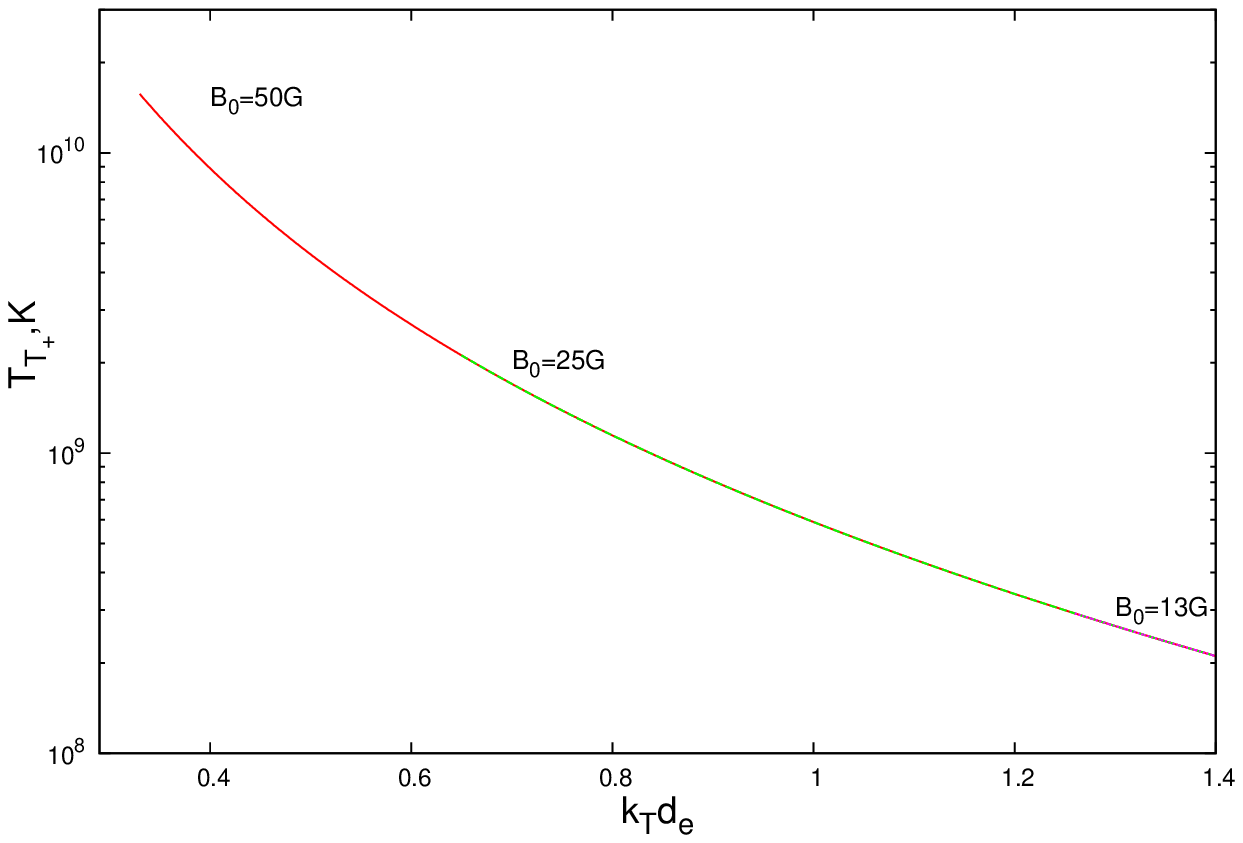} 
\includegraphics[angle=0, width=10cm]{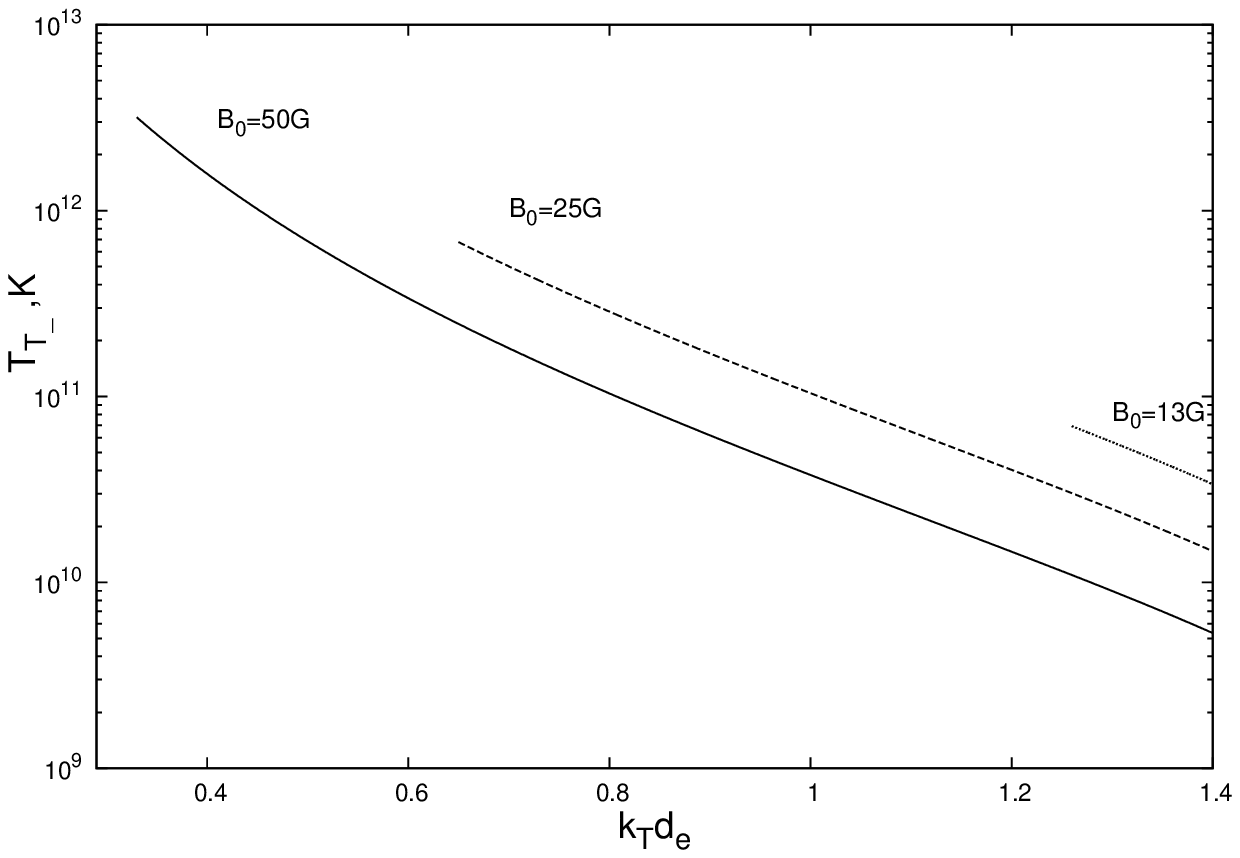} 
\caption{ Brightness temperature $T_T(k)$ of fundamental radio emission for the
f- process (top panel), and the d- process (bottom panel). Here we
consider a nonthermal level of LWs, thermal ISWs and $\protect\eta=4\protect%
\pi ^{2}$. The plasma parameters are $n_{0}=2.8\times 10^{8}$ cm$^{-3}$; $%
T_{e}=10^{5}$ K; $T_{i}/T_{e}=0.5$ and three values of magnetic field
strength $B_0 = 50$, $25$ and $13$ G.}
\label{fig:satfdLpT_5}
\end{figure*}

Figures \ref{fig:satfdLpT_6}--\ref{fig:satfdLpT_5} show the
brightness temperature $T_{T}(k)$ of fundamental radio emission for the 
f- process (top panel), and for the d- process (bottom panel) for two
sets of parameters: $T_e=10^6$ K and $B_0=150$--$40$ G and $T_e=10^5$ K
and $B_0=50$--$13$ G . For both cases the values of
brightness temperature $T_{T}(k)$ are rather high even for the thermal level
of ion-sound waves.

\subsubsection{Case ii): $W_{S}/\protect\omega _{S}\gg W_{L}/\protect\omega %
_{L}$}

In this case the f- and d- processes are essentially equivalent
and saturate at
\begin{equation}
T_{T+}=T_{T-}=\omega _{T}\frac{\eta }{k_{S}^{2}}\frac{W_{L}}{\omega _{L}}.
\label{dsatL}
\end{equation}%
It follows from Figures \ref{fig:satfdLpT_6}--\ref{fig:satfdLpT_5}
(bottom panel) that the brightness temperature of radio emission is in the
range: $T_{T\pm }\approx 2\times 10^{12}$--$10^{15}$ K for $T_{e}=10^{6}$ K
and $T_{T\pm }\approx 5\times 10^{9}$--$2\times 10^{13}$ K for $T_{e}=10^{5}
$ K.

Let us specify the non-thermal level of ion-sound waves, which is required
for this to be the case. The 1D spectral energy density of ISWs is related to
the spectral density of electron density fluctuation as (see Appendix C for
details)
\begin{equation}
\frac{W_{S}\left( k\right) }{n_{e}T_{e}}=\left( 1+k_{S}^{2}\lambda
_{De}^{2}\right) \frac{\left\vert \delta n_{e}\right\vert _{k}^{2}}{n_{e}^{2}%
}.  \label{deltan_k}
\end{equation}%
According to above expression the density spectrum of thermal level of ion
sound waves corresponds to:
\begin{equation}
\frac{\left\vert \delta n_{e}\right\vert _{thk}^{2}}{n_{e}^{2}}=\frac{4\pi }{%
n_{e}\lambda _{De}^{2}}\frac{k_{S}^{2}\lambda _{De}^{2}}{1+k_{S}^{2}\lambda
_{De}^{2}}.
\end{equation}

\begin{figure*}[h!]
\includegraphics[angle=0, width=10cm]{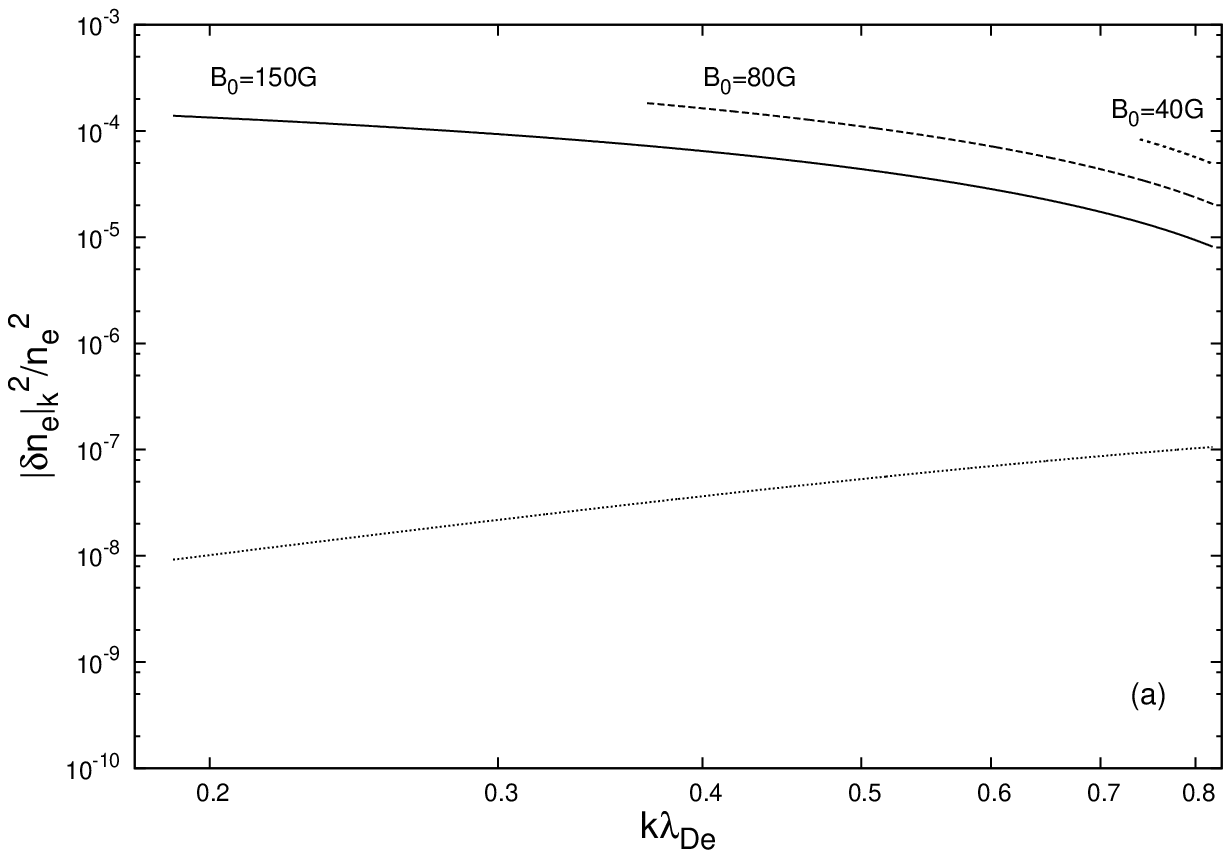} 
\includegraphics[angle=0, width=10cm]{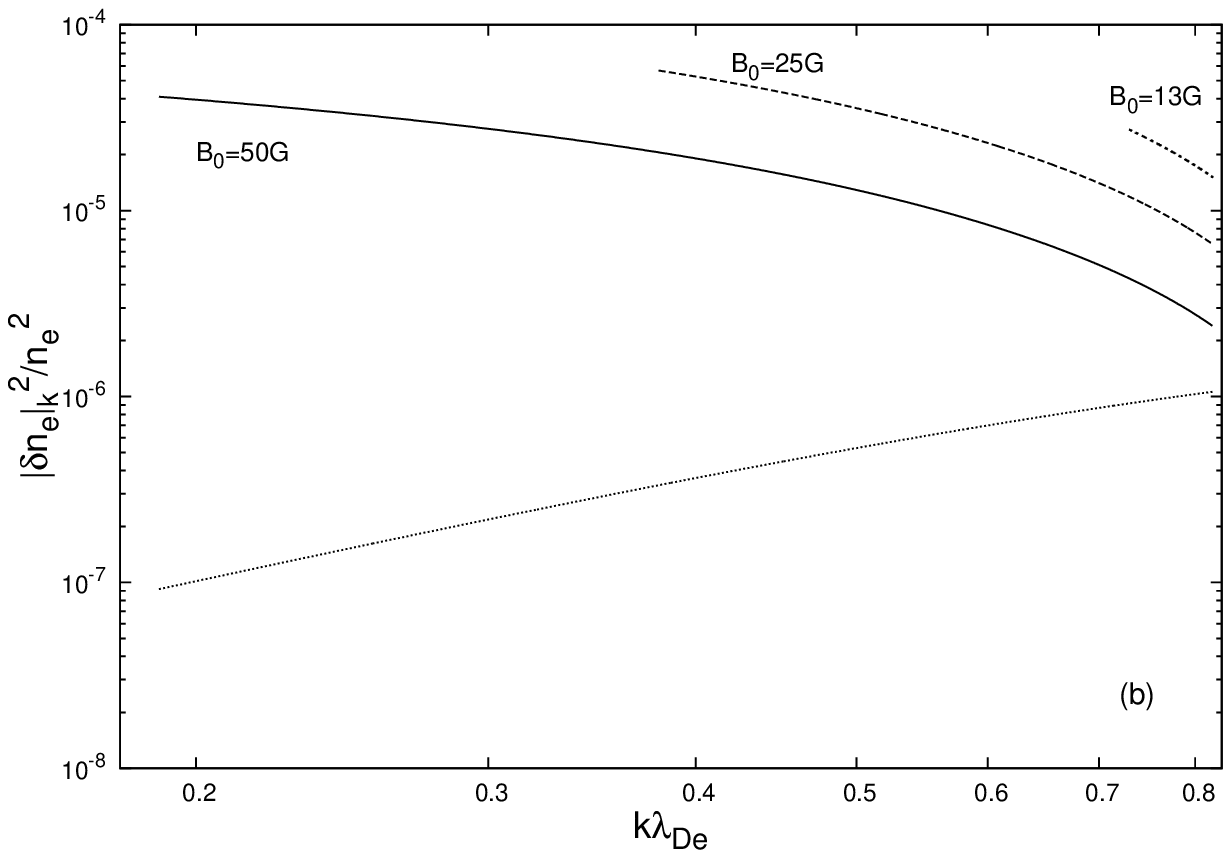}
\caption{Spectral density of electron density fluctuation versus $k\protect%
\lambda_{De}$ for: thermal level of ion-sound waves (dotted line); when $%
W_{k}^{s}/\protect\omega _{s}=W_{k}^{l}/\protect\omega _{l}$. Here we used
the following background plasma parameters $n_{0}=2.8\times 10^{8}$ cm$^{-3}$%
; $T_{i}/T_{e}=0.5$. Case (a) corresponds to $T_e=10^6$ K and three values
of magnetic field strength $B_{0}=150$ G (solid line), $B_{0}=80$ G (dashed
line), $B_{0}=40$ G (dashed-dotted line). Case (b) corresponds to $T_e=10^5$
K and three values of magnetic field strength $B_{0}=50$ G (solid line), $%
B_{0}=25$ G (dashed line), $B_{0}=13$ G (dashed-dotted line)}
\label{fig:deltank_n0_2plots}
\end{figure*}

In Figure \ref{fig:deltank_n0_2plots} we present the spectral
density of the electron density fluctuation for the thermal level of ion-sound
waves (dotted line) and when $W_{S}/\omega _{s}=W_{L}/\omega _{L}$ (solid,
dashed, dashed-dotted lines). Consequently, for $T_{e}=10^{6}$ K the case $%
W_{S}/\omega _{S}\gg W_{L}/\omega _{L}$ requires a level of the ISW that is
several orders of magnitude larger then the thermal level ($\left\vert
\delta n_{e}\right\vert _{k}^{2}/\left\vert \delta n_{e}\right\vert
_{thk}^{2}\gtrsim 10^{4}$ for $k\lambda _{De}\gtrsim 0.2$). For the lower
temperature $T_{e}=10^{5}$ K in the short wavelength domain, the amplitudes of the ISWs
approach the thermal level, but still remain larger then the thermal level.

\begin{figure*}[h!]
\includegraphics[angle=0, width=10cm]{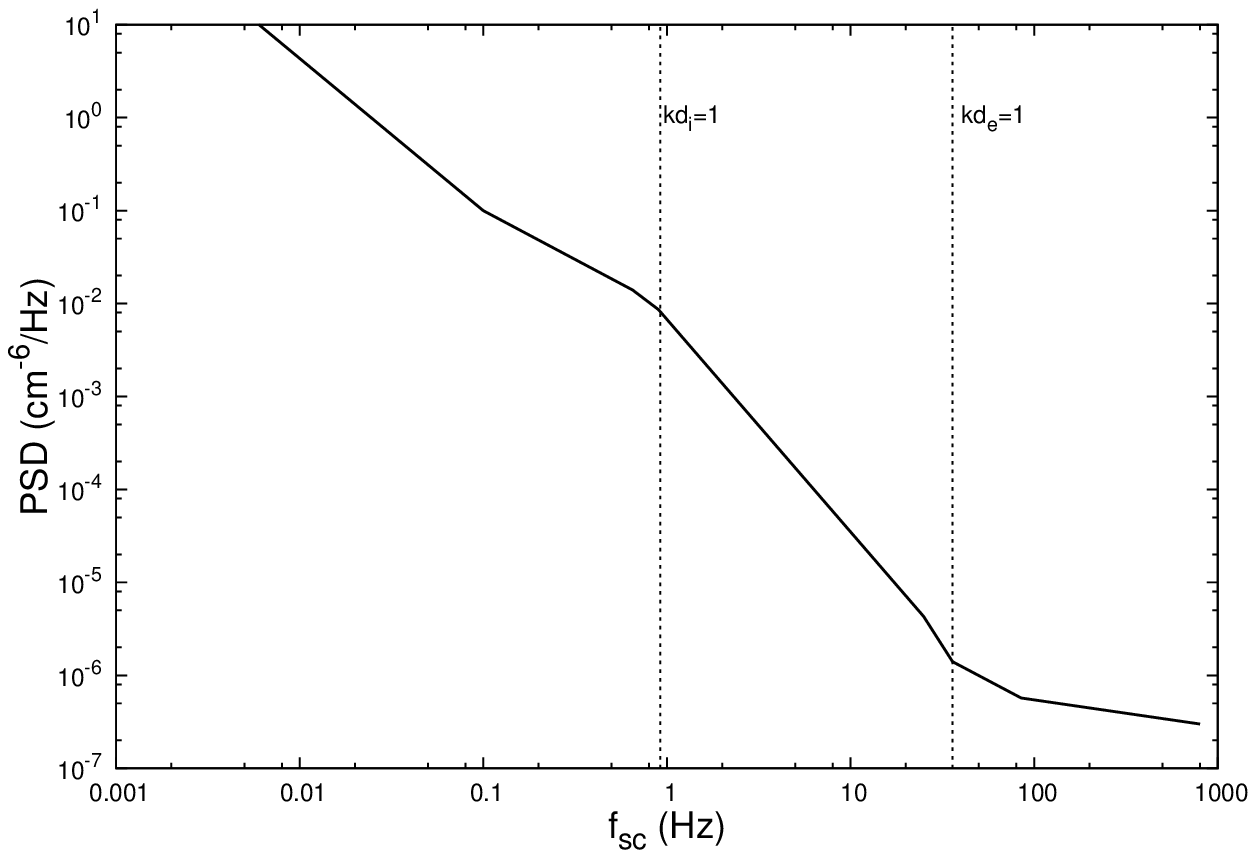} %
\includegraphics[angle=0, width=10cm]{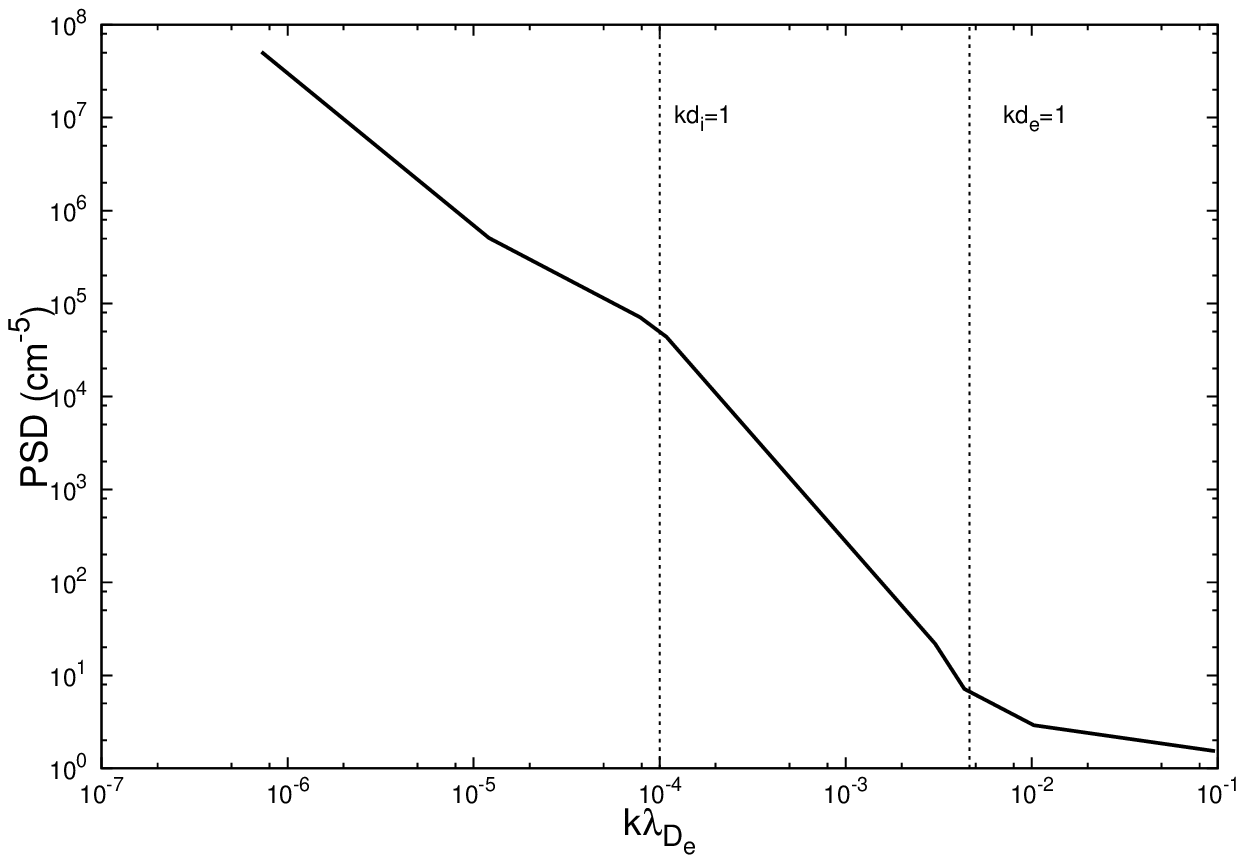}
\caption{\citet{2012PhRvL.109c5001C} frequency-- power spectra of solar wind
electron fluctuations (top panel), and $k$-- power spectra of solar wind electron
fluctuations (bottom panel). Here we used the following background solar
wind plasma parameters $V_{\mathrm{SW}}=3.2\times 10^{7}$ cm/s, $n_{i}=16$ cm$^{-3}$, $%
T_e=10^5$ K ($\protect\lambda_{De}=616$cm).}
\label{fig:Chen}
\end{figure*}

Our model describes emission from active regions in the corona, where the
relative level of density fluctuations is unknown. %
\citet{2012PhRvL.109c5001C} reported a measurement of the spectral index of
density fluctuations between ion and electron scales in solar wind
turbulence. In Figure \ref{fig:Chen} (top panel) the power spectra of
electron density fluctuations in the solar wind is shown according to the
Figure 2 of \citet{2012PhRvL.109c5001C}.

Assuming that the frequency--spectra are Doppler shifted $k$-- spectra
\[
P(k)=\frac{P(f)V}{2\pi },
\]%
with $k=2\pi f/V$ \ and $V\equiv V_{SW}=3.2\times 10^{7}$ cm/s we can get the
$k$-- power spectra of solar wind electron fluctuations Figure \ref{fig:Chen}
(bottom panel).

\begin{figure*}[tbp]
\includegraphics[angle=0, width=10cm]{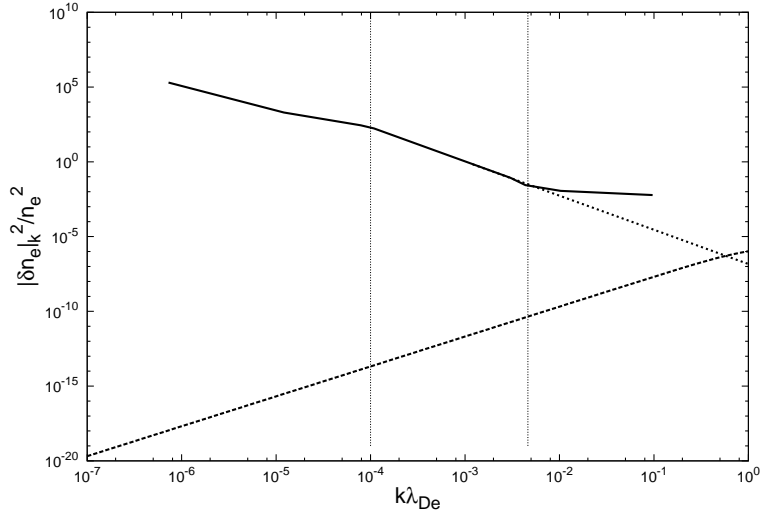}
\caption{Spectral density of electron density fluctuation versus $k\protect%
\lambda_{De}$ for: the thermal level of ion-sound waves (dashed line); the density
fluctuation spectrum of solar wind turbulence given by
\citet{2012PhRvL.109c5001C} (solid line). Here we used the following
background plasma parameters $V_{SW}=3.2\times 10^{7}$ cm/s, $n_{i}=16$ cm$%
^{-3}$, $T_e=10^5$ K ($\protect\lambda_{De}=616$cm) for solar wind.}
\label{fig:deltank_n0_obsEx}
\end{figure*}

We extrapolate the $k$-- power spectra of solar wind electron fluctuations
(see Figure \ref{fig:deltank_n0_obsEx}) into the short wavelength domain using $%
k^{-2.7}$ (dotted line) and present the spectral density of electron density
fluctuation for the thermal level of solar wind ion-sound waves (dashed
line).

In the relevant range $0.19<k\lambda_{De}<0.82$ the observed density
fluctuation is of the order of thermal fluctuations. We would like to notice that the observed $-2.7$ spectra are 
attributed to the KAW turbulence \citep{2012PhRvL.109c5001C} rather than to ISWs, however we consider this extrapolations as the upper bound. Also Helios spacecraft
observations at 0.3AU \citep{1990JGR....9511945M} and numerical models %
\citep{2010ApJ...721..864R} suggest that the relative level of density
fluctuations decreases toward the Sun. Consequently, it seems unlikely that
the case $W_{S}/\omega _{S}\gg W_{L}/\omega_{L}$ is realistic.

\subsection{Absorption of Emission During the Propagation}

\label{subS-absorption} Further, we have considered the collisional
absorption of radiation during propagation. Accordingly to %
\citet{2014A&A...562A..57R} collisional absorption (inverse bremsstrahlung)
with damping rate $\gamma _{d}$ gives an optical depth:
\begin{equation}
\tau =\int\limits_{0}^{1{\mathrm AU}}\frac{\gamma _{d}\left( x\right) }{%
V_{g}^{T}\left( x\right) }\rmd x,
\end{equation}%
where
\begin{equation}
\gamma _{d}=\gamma _{c}\frac{\omega _{pe}^{2}}{\omega ^{2}}=\frac{1}{3}\sqrt{%
\frac{2}{\pi }}\frac{\Gamma }{V_{Te}^{3}}\frac{\omega _{pe}^{2}}{\omega ^{2}}%
=\frac{1}{3}\sqrt{\frac{2}{\pi }}\frac{e^{2}\ln \Lambda }{m_{e}V_{Te}^{3}}%
\frac{\omega _{pe}^{4}}{\omega ^{2}},
\end{equation}%
and
\begin{equation}
V_{g}^{T}=\frac{\partial \left( \omega _{pe}^{2}+k_{T}^{2}c^{2}\right) }{%
\partial k_{T}}=\frac{c}{\omega }\left( \omega ^{2}-\omega _{pe}^{2}\right)
^{1/2}.
\end{equation}%
For emission at a frequency $\omega _{0}$ we then have
\begin{equation}
\tau =\frac{1}{3}\sqrt{\frac{2}{\pi }}\frac{e^{2}\ln \Lambda }{%
m_{e}V_{Te}^{3}}\frac{1}{c\omega _{0}}\int\limits_{0}^{1{\mathrm AU}}\frac{\omega
_{pe}^{4}\left( x\right) }{\left( \omega _{0}^{2}-\omega _{pe}^{2}\left(
x\right) \right) ^{1/2}}\rmd x.
\end{equation}%
Assuming an exponential density profile $n_{e}(x)=n_{0}\exp (-x/H)$, where $%
x $ is the distance from the region of emission at density $n_{0}$, and $H$
is the density scale height, we found the following (slighter corrected
compared with \citealp{2014A&A...562A..57R}) expression for the optical
depth
\begin{equation}
\tau =\frac{4}{9}\sqrt{\frac{2}{\pi }}\frac{e^{2}\ln \Lambda }{%
m_{e}V_{Te}^{3}}\frac{H}{c}\left\{ \omega _{0}^{2}-\frac{1}{\omega _{0}}%
\left[ \omega _{0}^{2}-\omega _{pe}^{2}\left( 0\right) \right] ^{1/2}\left[
\omega _{0}^{2}+0.5\omega _{pe}^{2}\left( 0\right) \right] \right\} .
\label{tau}
\end{equation}%
As we consider a dense coronal plasma we can use the reasonable value of $%
H=5\times 10^{9}$ cm and find that%
\begin{equation}
\tau \approx 2.8\left[ \overline{\omega }_{0}^{2}-\frac{1}{\overline{\omega }%
_{0}}\left( \overline{\omega }_{0}^{2}-1\right) ^{1/2}\left( \overline{%
\omega }_{0}^{2}+0.5\right) \right] ,
\end{equation}%
where $\overline{\omega }_{0}=\omega _{0}/\omega _{pe}\left( 0\right) .$

\begin{figure*}[tbp]
\includegraphics[width=7.3cm, angle=270]{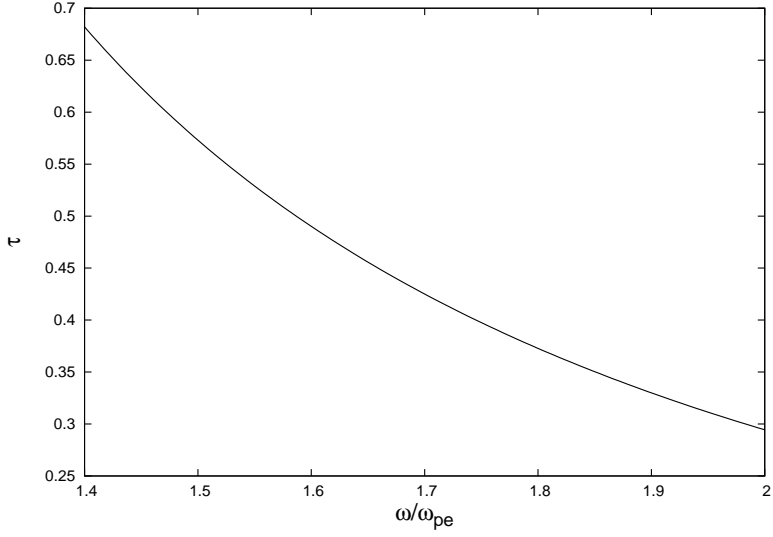}
\caption{The optical depth as a function of $\protect\omega /\protect\omega %
_{pe}$, for $H=5\times 10^{9}$ cm and $f_{pe}=\protect\omega _{pe}/2\protect%
\pi =0.2$ GHz}
\label{fig:tau}
\end{figure*}

As seen from Figure \ref{fig:tau} for such a density scale height the
resulting escape fraction is rather high for $f_{pe}=0.2$ GHz. If for a
different choice of parameters, we could increase $\tau $ by an order of
magnitude, we would predict observable emissions due to quite high values of
brightness temperature for the f- and d- processes.

\section{Application to Type I Solar Radio Bursts}

\label{S-Application}

Type I emission in Noise storms is the most commonly observed radio
phenomenon of the Sun (see review by
\citealp{1977sns..book.....E,
1985srph.book.....M}). Myriads of type I bursts, each lasting about a
second, are superimposed on a continuum, which lasts from a few hours to
days. Type I emission can extend from 50 to 500 MHz, peaking around
150--200 MHz. This activity occurs in active regions above sunspots, with
relatively strong magnetic fields. Type I emission is strongly ($\approx
100\%$) circularly polarized, and there is only fundamental emission with no
harmonic component. The brightness temperature is in the range from $10^{7}$
to $10^{10}$ K for continuum emissions and can exceed $10^{11}$ K for
bursts. The sizes of the emission sources are several arcminutes for the
continuum and about 1 arcminute for bursts.

Until now, the plasma emission mechanism proposed by %
\citet{1980SoPh...67..357M} is the most popular interpretation for type I
radio emission. The plasma emission mechanism consists of two steps: i)
isotropic or loss-cone distribution of energetic electrons generates LWs; ii)
these LWs generate electromagnetic radiation (radio waves) via nonlinear
coupling with ion-acoustic or other low-frequency waves. In such models, the
type I continuum is explained by the LWs\ with effective temperature $%
T_{L}\gtrsim $ $10^{9}$ K generated by a \textquotedblleft gap electron
distribution\textquotedblright. This implies a high level of low-frequency
waves to ensure that the emission mechanism saturates at the radio
brightness temperature $T_{T}$ equal to the Langmuir wave temperature $T_{L}$%
. The increasing $T_{T}$ during bursts is attributed to the local LW
enhancements driven by the loss-cone instability. Theories based on the
plasma emission mechanism have been developed in greater detail
\citep{1981A&A....94..100B,1982A&A...105..221S, 1986SoPh..103..141W,
1991SoPh..132..173T}.

One of the weak points of such theories is that the isotropic velocity
distribution of electrons should give rise to the harmonic emission as well,
which was never observed in type I emission. Also, a high level of
low-frequency waves is required in order that the emission mechanism
saturates at a brightness temperature equal to the plasma wave temperature $%
T_{L}$.

We propose an alternative model for type I radio emission. The
imbalanced turbulence of upward-propagating IAWs forms an asymmetric
electron velocity distribution with a quasilinear plateau in its forward
half at $\sqrt{1+T_{i}/T_{e}}V_{Te}<V_{\parallel }<V_{A}$. Landau damping of
LWs propagating in the same direction in this velocity range is reduced,
which makes possible the spontaneous growth of their amplitudes. In
previous sections, we calculated the nonthermal level of these LWs and showed
that it is sufficient to generate the observed type I radio emission. As the
strongest radio emission is generated by LWs with phase velocities $%
V_{\mathrm{LW}}\lesssim V_{A}$, the full plateau extending down to $\sqrt{%
1+T_{i}/T_{e}}V_{Te}$ is not required. In fact, it is sufficient that a
narrower plateau is formed in the vicinity of $V_{\parallel }\lesssim V_{A}$%
.

Our estimations show that even with the thermal-level ion-acoustic waves, we
get the observed brightness temperature for type I emission. The harmonic
emission is never observed in type I storms; it does not occur also in our
model, because the back-scattered LWs meet a strong Landau damping, $\gamma
_{L}\approx \omega _{pe}$. 

Our model works in low-beta plasmas, where the Alfv\'{e}n velocity
exceeds the electron thermal speed ($V_{A}>V_{Te}$). Such
conditions can be found in the solar corona at radial distances $\approx 1.1$
solar radii, where type I storms are generated. Recent
observations have demonstrated that magnetic fields above active regions at
these distances can be strong enough, several $10$s of G, and
temperatures low, down to $10^{4}$ K (see
examples in Figure 8 by \citealp{2016ApJ...833....5S} with instances of $B_{0}$ from 
$10$ to $150$ G, and paper by \citealp{2015ApJ...806...81A}
with coronal $T_{e}$ from $10^{4}$ to $10^{6}$ K). 

The often observed phenomenon of coronal rain provides a
well-documented example of cold plasma patches created at high coronal
levels by thermal instability (see \citealp{2015ApJ...806...81A}, and
references therein). Plasma in the upper parts of long magnetic loops is
especially prone to this instability. As the type I emission is generated,
tentatively, at the tops of the highest magnetic loops overlapping active
regions, the relatively cold plasma, produced there by thermal instability,
may provide a suitable conditions for type I radio emission.

We note that \citet{1978SoPh...57..279D} and \citet{1986BAICz..37..115G} proposed empirical models for the coronal $B$ based on radio observations. In the model by \citet{1978SoPh...57..279D}, the magnetic field strength depends on the plasma density model and assumes a particular generation mechanism for meter-wavelength radio bursts, with type I bursts excluded from the analysis. The model by \citet{1986BAICz..37..115G} is based on type I radio observations but assumes a particular generation mechanism for the bursts implying numerous shocks; it also critically depends on the radial profile of coronal density. It seems that the conditions above active regions are highly variable and can hardly be described in the framework of a single model. Therefore, instead of using a model, we refer to the values of coronal magnetic field measured directly by \citet{2016ApJ...833....5S}, which do not require assumptions of any specific plasma model or process. The measured values, several tens of gauss at relevant heliocentric distances $R/{\mathrm R_{\odot }}\approx  1.1$, appeared to be larger than the values predicted in previous models. With these measured values of $B$, conditions for our scenario can be easily satisfied. 

Additionally we have to note that current model reproduces the almost $100\%$ polarization in the o-mode of type I bursts. The main reason is that the frequency of the emission due to the
$L\pm S\rightarrow T$ processes is below the cutoff frequency of the extraordinary (x-) mode,
$\omega_{\mathrm x}=[\omega_{Be}+(\omega_{Be}^2+4\omega_{pe}^2)^{1/2}]/2$ for the values of the ratio, $\omega_{Be}/\omega_{pe}$, required in our model.
The resulting electromagnetic emission then must be $100\%$ in the o-mode.

\section{Conclusions}

\label{S-Conclusions} We investigated the influence of imbalanced
small-scale IAW turbulence on the spontaneous growth of LWs. The resulting
high-amplitude LWs can generate type I radio emission observed above
active regions in the solar corona.

Our starting point has been that the imbalanced turbulence of
forward-propagating IAWs forms an asymmetric quasilinear plateau in the
forward half of the electron velocity distribution. This leads to the
suppression of Landau damping for resonant LWs in the range of phase
velocities $\sqrt{1+T_{i}/T_{e}}V_{Te}<$ $V_{LW}<$ $V_{A}$. As a
consequence, spontaneous excitation of high-amplitude LWs with $%
W_{L}/W_{therm}\approx 10^{7}$--$10^{9}$ (for $T_{e}=10^{6}$ K) and
$W_{L}/W_{therm}\approx 10^{5}$--$10^{7}$ (for $T_{e}=10^{5}$ K) occurs in
this velocity range. These LWs can produce strong electromagnetic radiation
at the fundamental frequency close to the electron plasma frequency.

Even with the unfavorable thermal level of ion-sound waves, the brightness
temperature of radio emission in our model is $T_{T+}\approx
10^{9}$--$10^{11}$ K, $T_{T-}\approx 10^{12}$--$10^{15}$ K (for $%
T_{e}=10^{6}$ K) and $T_{T+}\approx 10^{8}$--$10^{10}$ K, $T_{T-}\approx
5\times 10^{9}$--$10^{12}$ K (for $T_{e}=10^{5}$ K), which is high enough to
explain observations. The bursts with extremely high brightness temperatures
$\gtrsim $ $10^{11}$ K can be easier generated by the d- process. 
Moreover our theory predicts $100\%$ polarization in the o-mode of type I emission.

In conclusion, our model with the imbalanced IAW turbulence and
spontaneously excited LWs provides a feasible explanation for solar type
I radio emission. This model is also consistent with the fact that the first
harmonic is never observed in type I radio emission. 

Finally, we note that the 3D velocity distribution of electrons leads to a
relativistic correction to the Landau damping of Langmuir waves %
\citep{1977AuJPh..30..481M} that may exceed the collisional damping assumed
here. This effect applies to an isotropic electron distribution with a gap
over some velocity range. The generalization of this gap model to the
plateau model assumed here has yet to be investigated.

\begin{acks}
The authors are thankful to the anonymous referee and to the Guest Editor Alexander Nindos for constructive comments. E. P. K. and N. H. B. were supported by Science and Technology Facilities Council Grant No. ST/L000741/1. Y.V. was supported by the Belgian Science Policy Office through IAP Programme, project P7/08 CHARM.
\end{acks}

\begin{acks}[Disclosure of Potential Conflicts of Interest]
 The authors declare that they have no conflicts of interest.
\end{acks}

\appendix

\section{The Parallel Electric Field of Dispersive Alfv\'{e}n Waves}

Taking as in \citet{2011A&A...527A.130B}, the first moments of the
linearized drift-kinetic equation for the electrons, yields the linearized
continuity equation
\begin{equation}
\frac{\partial n_{e}}{\partial t}+\nabla_{\parallel}n_{0}u_{\parallel e}=0,
\end{equation}
where $n_{e}/n_{0}$ is the electron density perturbation relative to a
constant background $n_{0}$, $u_{\parallel e}$ is the parallel (to the
magnetic field) component of the electron fluid velocity and $%
\nabla_\parallel$ denotes the spatial gradient along the magnetic field. The
linearized parallel electron momentum equation is
\begin{equation}
n_{0}m_{e}\frac{\partial u_{\parallel e}}{\partial t}=-T_{e}\nabla_{%
\parallel}n_{e}-n_{0}eE_{\parallel},
\end{equation}
where we have used $p_{e}=n_{e}T_{e}$ for the pressure perturbation $p_{e}$
and assumed a constant electron temperature $T_{e}$. The parallel component
of the electric field $E_{\parallel}$ is related to the electrostatic
potential $\phi$ and the parallel component of the magnetic vector potential
$A_{\parallel}$ via Faraday's law,
\begin{equation}
E_\parallel=-\nabla_{\parallel}\phi -\frac{1}{c}\frac{\partial A_{\parallel}%
}{\partial t}.
\end{equation}
The parallel component of Ampere's law is
\begin{equation}
\nabla_{\perp}^{2}A_{\parallel}=\frac{4\pi}{c}J_{\parallel}
\end{equation}
where $\nabla_\perp$ is the spatial gradient perpendicular to the field and
it is assumed that the parallel current,
\begin{equation}
J_{\parallel}=-en_{0}u_{e\parallel},
\end{equation}
is carried only by electrons. The system is closed by the quasineutrality
condition
\begin{equation}
n_{e}=(\Gamma_{0}-1)\frac{en_{0}}{T_{i}}\phi,
\end{equation}
where $\Gamma_{0}$ is an integral operator which describes the average of
the electrostatic potential over a ring of Larmor radius $\rho_{i}$. In
Fourier space, $\Gamma(b)=e^{-b}I_{0}(b)$ where $b=\rho_{i}^{2}k^{2}_{\perp}$
and $I_{0}$ is the modified Bessel function. We will use a simple Pade
approximant for the operator $\Gamma_{0}$ given by $\Gamma_{0}(b)-1=-b/(1+b)$%
. Adopting the following MHD normalization
\begin{equation}
(\hat{t},\hat{\nabla}_{\parallel\perp},\hat{A}_{\parallel},\hat{\phi},\hat{E}%
_{\parallel})=(\frac{t}{\tau_{A}},\nabla_{\parallel\perp}L ,\frac{%
A_{\parallel}}{LB_{0}},\frac{\phi c}{LV_{A}B_{0}},\frac{E_{\parallel}c}{%
V_{A}B_{0}}),
\end{equation}
with $\tau_A= L/V_A$, we therefore obtain
\begin{equation}
\partial_{t} n_{e}-\nabla_{\parallel}J_{\parallel}=0,  \label{A8}
\end{equation}
\begin{equation}
\partial_{t}(\psi-d_{e}^{2}J_{\parallel})+\nabla_{\parallel}(%
\rho_{s}^{2}n_{e}-\phi)=0,  \label{A9}
\end{equation}
\begin{equation}
J_{\parallel}=\nabla^{2}_{\perp}\psi,  \label{A10}
\end{equation}
\begin{equation}
(1-\rho_{i}^{2}\nabla_{\perp}^{2})n_{e}=\nabla_{\perp}^{2}\phi,  \label{A11}
\end{equation}
with $\psi=-A_{\parallel}$. This model generalizes the one already discussed
in \citet{2010PhPl...17f2308B} to include the effect of electron inertia,
the term proportional to $d_{e}^{2}$ in Equation (\ref{A9}), which can be
rewritten as
\begin{equation}
E_{\parallel}=-\rho_{s}^{2}\nabla_{\parallel}n_{e}+d_{e}^{2}\partial_{t}J_{%
\parallel}.
\end{equation}
This is Ohm's law (the parallel electron momentum equation) and there are
two contributions to the parallel electric field. The first is produced by
electron density fluctuations along the magnetic field and involves the
ion-sound Larmor radius $\rho_{s}$ (the ion gyroradius at the electron
temperature, {\ie}  $\rho^{2}_{s}=(T_{e}/T_{i})\rho^{2}_{i}$). The second is
produced by electron inertia and involves the electron skin depth $d_{e}$.
The Alfven wave equation is easily obtained from Equations (\ref{A8})--(\ref%
{A11}) and reads
\begin{equation}
\partial_{tt}(\psi-d_{e}^{2}\nabla_{\perp}^{2}\psi)=\nabla_{%
\parallel}^{2}[1-(\rho_{i}^{2}+\rho_{s}^{2})]\nabla_{\perp}^{2}\psi,
\end{equation}
and hence the dispersion relation is
\begin{equation}
\omega=\pm k_{A\parallel}\sqrt{\frac{1+k^{2}_{A\perp}\rho^{2}_{T}}{%
1+k_{A\perp}^{2}d_{e}^{2}}},
\end{equation}
with $\rho_{T}^{2}=\rho_{i}^{2}+\rho_{s}^{2}$. We notice that when $%
\rho_{T}^{2}=d_{e}^{2}$, the dispersion relation is $\omega=\pm
k_{A\parallel}$: the wave becomes non-dispersive and the wave-particle
resonance reduces to a single point in velocity-space. Using the following
conductivity relation
\begin{equation}
\partial_{t}E_{\parallel}=(d_{e}^{2}\partial_{tt}-\rho_{s}^{2}\nabla^{2}_{%
\parallel})J_{\parallel},
\end{equation}
we observe that the parallel electric field remains finite in the
non-dispersive regime, {\ie} $|E_{\parallel}|=\rho_{i}^{2}k_{A\parallel}k_{A%
\perp}|\delta B_{\perp}|$ in this case, its amplitude is proportional to
the ion temperature. This property should be contrasted with Hall-MHD (see %
\citealp{2009PhPl...16f4503B} and references therein) which is based on a cold ion
assumption and therefore lacks a parallel electric field in the parameter
regime $\rho_{s}=d_{e}$.

\section{An Angle-averaged Model for Fundamental Radio Emission}

The kinetic Equation (\ref{kin_eq}) for the processes $L\pm S\rightarrow T$
may be written in the form
\begin{eqnarray}
&&\frac{\rmd W_{T\pm}\left( \vec{k}_{T}\right) }{\rmd t}=\pi \frac{\omega
_{pe}^{3}}{4n_{e}T_{e}}\left( 1+\frac{3T_{i}}{T_{e}}\right) \int \rmd
\vec{k}_{L} \rmd \vec{k}_{S}\times \\
&&\omega _{S}\frac{\left[ \vec{k}_{T}\times \vec{k}_{L}\right] ^{2}}{%
k_{T}^{2}k_{L}^{2}}\left[ \frac{W_{L}}{\omega _{L}}\left( \frac{W_{S}}{%
\omega _{S}}\mp \frac{W_{T}}{\omega _{T}}\right) -\frac{W_{T}}{\omega _{T}}%
\frac{W_{S}}{\omega _{S}}\right] \times  \nonumber \\
&&\delta \left( \vec{k}_{T}-\vec{k}_{L}\mp \vec{k}_{S}\right)
\delta \left( \omega _{T}-\omega _{L}\mp \omega _{S}\right) ,  \nonumber
\end{eqnarray}%
where $W_L$, $W_S$ and $W_T$ are functions of their respective wavenumbers.
First we perform the integral over $\mathbf{k}_{S}$ using $\delta \left(
\mathbf{k}_{T}-\mathbf{k}_{L}\mp \mathbf{k}_{S}\right) $ to get
\begin{eqnarray}
&&\frac{\rmd W_{T\pm}\left( \vec{k}_{T}\right) }{\rmd t}=\pi \frac{\omega
_{pe}^{3}}{4n_{e}T_{e}}\left( 1+\frac{3T_{i}}{T_{e}}\right)\int \rmd \vec{%
k}_{L}\frac{\left[ \vec{k}_{T}\times \vec{k}_{L}\right] ^{2}}{%
k_{T}^{2}k_{L}^{2}}\omega _{S}\left( \vec{k}_{S}\right) \times \\
&&\left[ \frac{W_{L}\left( \vec{k}_{L}\right) }{\omega _{L}\left( \vec{%
k}_{L}\right) }\left( \frac{W_{S}\left( \vec{k}_{S}\right) }{\omega
_{S}\left( \vec{k}_{S}\right) }\mp \frac{W_{T}\left( \vec{k}%
_{T}\right) }{\omega _{T}\left( \vec{k}_{T}\right) }\right) -\frac{%
W_{T}\left( \vec{k}_{T}\right) }{\omega _{T}\left( \vec{k}_{T}\right) }%
\frac{W_{S}\left( \vec{k}_{S}\right) }{\omega _{S}\left( \vec{k}%
_{S}\right) }\right] \times  \nonumber \\
&&\delta \left( \omega _{T}\left( \vec{k}_{T}\right) -\omega _{L}\left(
\vec{k}_{L}\right) \mp \omega _{S}\left( \vec{k}_{S}\right) \right) ,
\nonumber
\end{eqnarray}%
Here $\vec{k}_{S}=\pm \vec{k}_{T}\mp \vec{k}_{L},$ for the f- and d- processes, respectively.

Assuming that the LWs and ion sound waves have some small angular spread in
wavenumber space, covering a solid angle of $\Delta \Omega $, and further
assuming that they are uniform within this spread, fundamental emission is
produced approximately isotropically. So we define:
\begin{eqnarray}
W_{L,S}\left( \vec{k}\right) &=&\frac{1}{\Delta \Omega k^{2}}%
W_{L,S}^{Av}\left( k\right) ,  \label{def_W}
\end{eqnarray}
within $\Delta \Omega$ the small solid angle occupied by the parent waves,
and zero elsewhere, with $W_{L,S}^{Av}\left( k\right)$ defined by
\begin{eqnarray}
W_{L,S}^{Av}\left( k\right) &=&\int \int k^{2}\sin \theta W_{L,S}\left(
\vec{k}\right) \rmd\theta\rmd\Phi,  \label{def_W_2nd}
\end{eqnarray}%
and consider isotropic transverse waves
\begin{eqnarray}
W_{T}\left( \vec{k}_{T}\right) &=&\frac{1}{4\pi k_{T}^{2}}%
W_{T}^{Av}\left( k_{T}\right),  \label{def_T} \\
W_{T}^{Av}\left( k\right) &=&\int \int k^{2}\sin \theta W_{T}\left( \vec{k%
}\right) \rmd\theta \rmd\Phi .  \label{def_T_2nd}
\end{eqnarray}

Writing $\rmd\vec{k}_{L}=k_{L}^{2}\sin \theta \rmd k_{L}d\theta \rmd\Phi $ and
substituting our definitions of the angle-averaged spectral energy densities
(expressions (\ref{def_W})--(\ref{def_W_2nd})), for the first term in the square
brackets we find:
\begin{eqnarray}
&&\int \int \sin ^{2}\theta _{LT}\frac{W_{L}\left( \vec{k}_{L}\right)
W_{S}\left( \vec{k}_{S}\right) }{\omega _{L}\omega _{S}}k_{L}^{2}\sin
\theta \rmd\theta\rmd\Phi \\
&=&\frac{W_{L}^{Av}\left( k_{L}\right) }{\Delta \Omega \omega _{L}}\frac{%
W_{S}^{Av}\left( k_{S}\right) }{\Delta \Omega k_{S}^{2}\omega _{S}}\int \int
\sin ^{2}\theta _{LT}\sin \theta \rmd\theta\rmd\Phi  \nonumber \\
&=&\frac{W_{L}^{Av}\left( k_{L}\right) }{\Delta \Omega \omega _{L}}\frac{%
W_{S}^{Av}\left( k_{S}\right) }{k_{S}^{2}\omega _{S}}\left\langle \sin
^{2}\theta _{LT}\right\rangle,  \nonumber
\end{eqnarray}%
where we assume the average value of $\sin^{2}\theta _{LT}$ is well defined
and given by
\begin{equation}
\int \int \sin ^{2}\theta _{LT}\sin \theta \rmd\theta\rmd\Phi=\Delta
\Omega\left\langle \sin^{2}\theta _{LT}\right\rangle.
\end{equation}

Similarly, using expressions (\ref{def_W})--(\ref{def_T_2nd}) for the second
and third terms in square brackets we write
\begin{equation}
\int \int \sin ^{2}\theta _{LT}\frac{W_{L}\left( \vec{k}_{L}\right)
W_{T}\left( \vec{k}_{T}\right) }{\omega _{L}\omega _{T}}k_{L}^{2}\sin
\theta {\mathrm d}\theta{\mathrm d}\Phi =\frac{W_{L}^{Av}\left( k_{L}\right) }{\omega _{L}}%
\frac{W_{T}^{Av}\left( k_{T}\right) }{4\pi k_{T}^{2}\omega _{T}}\left\langle
\sin ^{2}\theta _{LT}\right\rangle  \nonumber
\end{equation}
\[
\int \int \sin ^{2}\theta _{LT}\frac{W_{T}\left( \mathbf{k}_{T}\right)
W_{S}\left( \vec{k}_{S}\right) }{\omega _{T}\omega _{S}}k_{L}^{2}\sin
\theta {\mathrm d}\theta{\mathrm d}\Phi =\frac{W_{T}^{Av}\left( k_{T}\right) }{4\pi
k_{T}^{2}\omega _{T}}\frac{W_{S}^{Av}\left( k_{S}\right) }{\omega _{S}}%
\left\langle \sin ^{2}\theta _{LT}\right\rangle .
\]%
Finally the complete expression is:
\begin{eqnarray}
&&\frac{\rmd W_{T\pm}^{Av}\left( k_{T}\right) }{\rmd t}=\pi \frac{\omega _{pe}^{3}}{%
4n_{e}T_{e}}\left( 1+\frac{3T_{i}}{T_{e}}\right) \left\langle \sin
^{2}\theta _{LT}\right\rangle \times \\
&&\int \rmd k_{L}\omega _{S}\delta \left( \omega _{T}-\omega _{L}\mp \omega
_{S}\right) \times  \nonumber \\
&&\left[ \frac{W_{L}^{Av}\left( k_{L}\right) }{\omega _{L}}\frac{4\pi
k_{T}^{2}}{\Delta \Omega k_{S}^{2}}\frac{W_{S}^{Av}\left( k_{S}\right) }{%
\omega _{S}}\mp \frac{W_{L}^{Av}\left( k_{L}\right) }{\omega _{L}}\frac{%
W_{T}^{Av}\left( k_{T}\right) }{\omega _{T}}-\frac{W_{T}^{Av}\left(
k_{T}\right) }{\omega _{T}}\frac{W_{S}^{Av}\left( k_{S}\right) }{\omega _{S}}%
\right] ,  \nonumber
\end{eqnarray}%
where $k_{S}^{2}=k_{T}^{2}+k_{L}^{2}.$

Now, using $\delta \left( \omega _{T}-\omega _{L}\mp \omega _{S}\right) $ to
integrate over $k_{L}$, for $k_{T}^{2}d_{e}^2\gg \frac{1}{3}\frac{m_{e}}{%
m_{i}}$ we get
\begin{eqnarray}
&&\frac{\rmd W_{T\pm}^{Av}\left( k_{T}\right) }{\rmd t}=\frac{\pi \omega
_{pe}^{4}V_{s}}{24n_{e}T_{e}V_{Te}^{2}}\left( 1+\frac{3T_{i}}{T_{e}}\right)
\times \\
&&\left[ \frac{W_{L}^{Av}\left( k_{L}\right) }{\omega _{L}}\frac{4\pi
k_{T}^{2}}{\Delta \Omega k_{S}^{2}}\frac{W_{S}^{Av}\left( k_{S}\right) }{%
\omega _{S}}\mp \frac{W_{L}^{Av}\left( k_{L}\right) }{\omega _{L}}\frac{%
W_{T}^{Av}\left( k_{T}\right) }{\omega _{T}}-\frac{W_{T}^{Av}\left(
k_{T}\right) }{\omega _{T}}\frac{W_{S}^{Av}\left( k_{S}\right) }{\omega _{S}}%
\right] .  \nonumber
\end{eqnarray}%
Here $k_{L}\approx \mp \frac{k_{T}d_e}{\sqrt{3}\lambda _{De}}$ and we have
evaluated the average $\left\langle \sin ^{2}\theta _{LT}\right\rangle $
over a sphere, which gives a value of 1/2.

\section{Spectral Density of Electron Density Fluctuation due to Ion-Sound
Waves}

The relationship between the electric field and density perturbations due to
ion-sound waves is
\begin{equation}
\delta \vec{E=-}\frac{T_{e}}{n_{e}e}\mathbf{\nabla }\delta n.
\end{equation}%
We can rewrite the above expression as
\begin{equation}
\frac{\left\vert \vec{E}_{\vec{k}}\right\vert ^{2}}{4\pi }=\frac{1}{%
4\pi }\left( \frac{T_{e}}{n_{e}e}\right) ^{2}k_{s}^{2}\left\vert \delta
n_{e}\right\vert _{\vec{k}}^{2}.  \label{E_kA}
\end{equation}%
According to \citet{1995lnlp.book.....T} the occupation number $N_{\vec{k}%
}$ is related to $\vec{E}_{\vec{k}}$ as
\begin{equation}
N_{\vec{k}}^{l}=\left. \frac{\pi ^{2}}{\hbar }\left( \frac{1}{\omega ^{2}}%
\frac{\partial }{\partial \omega }\omega ^{2}\varepsilon ^{l}\right)
\right\vert _{\omega =\Omega ^{l}\left( \vec{k}\right) }\left\vert
\vec{E}_{\vec{k}}^{l}\right\vert ^{2},
\end{equation}%
where index $l$ denotes longitudinal waves. We can rewrite this expression
using that $\varepsilon ^{l}\left( \Omega ^{l},\vec{k}\right) =0$ and
\begin{equation}
N_{\vec{k}}^{l}=\frac{\left( 2\pi \right) ^{3}W_{\vec{k}}^{l}}{\hbar
\Omega ^{l}\left( \vec{k}\right) },
\end{equation}%
as
\begin{equation}
W_{\vec{k}}^{l}=\frac{1}{8\pi }\left. \omega \frac{\partial \varepsilon
^{l}}{\partial \omega }\right\vert _{\omega =\Omega ^{l}\left( \vec{k}%
\right) }\left\vert \vec{E}_{\vec{k}}^{l}\right\vert ^{2}.
\label{W_veckAp}
\end{equation}%
For ion-sound waves
\begin{equation}
\varepsilon ^{s}\approx 1-\frac{\omega _{pi}^{2}}{\omega _{s}^{2}}+\frac{1}{%
\left( k_{s}\lambda _{De}\right) ^{2}},  \label{eps_s}
\end{equation}%
which give the following dispersion relations for ion-sound waves
\begin{equation}
\omega _{s}\simeq \frac{k_{s}V_{s}}{\sqrt{1+k_{s}^{2}\lambda _{De}^{2}}}.
\label{disp_s}
\end{equation}%
According to expression (\ref{eps_s})
\begin{equation}
\frac{\partial \varepsilon ^{s}}{\partial \omega _{s}}=2\frac{\omega
_{pi}^{2}}{\omega _{s}^{3}}.  \label{part_eps}
\end{equation}%
Combining Equations (\ref{part_eps}) and (\ref{W_veckAp}) gives:
\begin{equation}
W_{\vec{k}}^{s}=\frac{\omega _{pi}^{2}}{\omega _{s}^{2}}\frac{\left\vert
\vec{E}_{\vec{k}}^{s}\right\vert ^{2}}{4\pi },
\end{equation}%
or using the dispersion relation (\ref{disp_s}) we can rewrite it as
\begin{equation}
W_{\vec{k}}^{s}=\frac{\left( 1+k_{s}^{2}\lambda _{De}^{2}\right) }{%
k_{s}^{2}\lambda _{De}^{2}}\frac{\left\vert \vec{E}_{\vec{k}%
}^{s}\right\vert ^{2}}{4\pi }.  \label{Wveck_nveck}
\end{equation}%
Inserting this expression in Equation (\ref{E_kA}) and noting that $\lambda
_{De}^{2}=T_{e}/(4\pi n_{e}e^{2})$, we get that the spectral energy density
of ISWs is related to the spectral density of electron density fluctuation
as
\begin{equation}
\frac{W_{\vec{k}}^{s}}{n_{e}T_{e}}=\left( 1+k_{s}^{2}\lambda
_{De}^{2}\right) \frac{\left\vert \delta n_{e}\right\vert _{\vec{k}}^{2}}{%
n_{e}^{2}}.  \label{Wvecknveck_app}
\end{equation}%
Here the following normalization is used
\begin{eqnarray}
\int \left( \delta n_{e}^{s}\right) _{\vec{k}}^{2}d\vec{k} &\vec{=}%
&\left( \delta n_e\right) ^{2}, \\
\int W_{\vec{k}}^{s}d\vec{k} &\mathbf{=}&W^{s}.
\end{eqnarray}%
Using that
\begin{equation}
\left( \delta n_e\right) _{\vec{k}}^{2}=\frac{\left( \delta n_e\right)
_{k}^{2}}{4\pi k_{s}^{2}},
\end{equation}%
and
\begin{equation}
W_{\vec{k}}^{s}=\frac{W_{k}^{s}}{4\pi k_{s}^{2}},
\end{equation}%
we can rewrite Equation (\ref{Wvecknveck_app}) for the one-dimensional case:
\begin{equation}
\frac{W_{k}^{s}}{n_{e}T_{e}}=\left( 1+k_{s}^{2}\lambda _{De}^{2}\right)
\frac{\left\vert \delta n_e\right\vert _{k}^{2}}{n_{e}^{2}}.
\label{deltan_kA}
\end{equation}

\bibliographystyle{spr-mp-sola}
\bibliography{Lyubchyk_etal}

\begin{thebibliography}{47}
\ifx\bisbn     \undefined \def\bisbn  #1{ISBN #1}\fi
\ifx\binits    \undefined \def\binits#1{#1}\fi
\ifx\bauthor   \undefined \def\bauthor#1{#1}\fi
\ifx\batitle   \undefined \def\batitle#1{#1}\fi
\ifx\bjtitle   \undefined \def\bjtitle#1{\textit{#1}}\fi
\ifx\bvolume   \undefined \def\bvolume#1{\textbf{#1}}\fi
\ifx\byear     \undefined \def\byear#1{#1}\fi
\ifx\bissue    \undefined \def\bissue#1{#1}\fi
\ifx\bfpage    \undefined \def\bfpage#1{#1}\fi
\ifx\blpage    \undefined \def\blpage #1{#1}\fi
\ifx\burl      \undefined \def\burl#1{\textsf{#1}}\fi
\ifx\href      \undefined \def\href#1#2{\textsf{#2}}\fi
\ifx\betal     \undefined \def\betal{\textit{et al.}}\fi
\ifx\bctitle   \undefined \def\bctitle#1{#1}\fi
\ifx\beditor   \undefined \def\beditor#1{#1}\fi
\ifx\bbtitle   \undefined \def\bbtitle#1{\textit{#1}}\fi
\ifx\bedition  \undefined \def\bedition#1{#1}\fi
\ifx\bseriesno \undefined \def\bseriesno#1{\textbf{#1}}\fi
\ifx\blocation \undefined \def\blocation#1{#1}\fi
\ifx\bsertitle \undefined \def\bsertitle#1{\textit{#1}}\fi
\ifx\bsnm      \undefined \def\bsnm#1{#1}\fi
\ifx\bsuffix   \undefined \def\bsuffix#1{#1}\fi
\ifx\bparticle \undefined \def\bparticle#1{#1}\fi
\ifx\barticle  \undefined \def\barticle#1{}\fi
\ifx\binstitute  \undefined \def\binstitute#1{#1}\fi
\ifx\bpublisher  \undefined \def\bpublisher#1{#1}\fi
\ifx\doiurl    \undefined
  \def\doiurl#1{\href{http://dx.doi.org/#1}{\textsf{DOI}}}\fi
\ifx\arxivurl  \undefined
  \def\arxivurl#1{\href{http://arxiv.org/abs/#1}{\textsf{arXiv}}}\fi
\ifx\adsurl    \undefined
  \def\adsurl#1{\href{http://adsabs.harvard.edu/abs/#1}{\textsf{ADS}}}\fi
\ifx\botherref \undefined \def\botherref#1{}\fi
\ifx\url       \undefined \def\url#1{\textsf{#1}}\fi
\ifx\bchapter  \undefined \def\bchapter#1{}\fi
\ifx\bbook     \undefined \def\bbook#1{}\fi
\ifx\bcomment  \undefined \def\bcomment#1{#1}\fi
\ifx\oauthor   \undefined \def\oauthor#1{#1}\fi
\ifx\citeauthoryear \undefined\def \citeauthoryear#1{#1}\fi
\ifx\endbibitem\undefined \def\endbibitem{}\fi
\ifx\bconflocation  \undefined \def\bconflocation#1{#1} \fi

\bibitem[\protect\citeauthoryear{{Antolin}
  \textit{et~al.}}{2015}]{2015ApJ...806...81A}
\begin{barticle}
\bauthor{\bsnm{{Antolin}}, \binits{P.}},
\bauthor{\bsnm{{Vissers}}, \binits{G.}},
\bauthor{\bsnm{{Pereira}}, \binits{T.M.D.}},
\bauthor{\bsnm{{Rouppe van der Voort}}, \binits{L.}},
\bauthor{\bsnm{{Scullion}}, \binits{E.}}:
\byear{2015},
\batitle{{The Multithermal and Multi-stranded Nature of Coronal Rain}}.
\bjtitle{\apj}
\bvolume{806},
\bfpage{81}.
\doiurl{10.1088/0004-637X/806/1/81}.
\adsurl{2015ApJ...806...81A}.
\end{barticle}
\endbibitem

\bibitem[\protect\citeauthoryear{{Artemyev}, {Zimovets}, and
  {Rankin}}{2016}]{2016A&A...589A.101A}
\begin{barticle}
\bauthor{\bsnm{{Artemyev}}, \binits{A.V.}},
\bauthor{\bsnm{{Zimovets}}, \binits{I.V.}},
\bauthor{\bsnm{{Rankin}}, \binits{R.}}:
\byear{2016},
\batitle{{Electron trapping and acceleration by kinetic Alfv{\'e}n waves in
  solar flares}}.
\bjtitle{\aap}
\bvolume{589},
\bfpage{A101}.
\doiurl{10.1051/0004-6361/201527617}.
\adsurl{2016A\%26A...589A.101A}.
\end{barticle}
\endbibitem

\bibitem[\protect\citeauthoryear{{Banerjee}
  \textit{et~al.}}{1998}]{1998A&A...339..208B}
\begin{barticle}
\bauthor{\bsnm{{Banerjee}}, \binits{D.}},
\bauthor{\bsnm{{Teriaca}}, \binits{L.}},
\bauthor{\bsnm{{Doyle}}, \binits{J.G.}},
\bauthor{\bsnm{{Wilhelm}}, \binits{K.}}:
\byear{1998},
\batitle{{Broadening of SI VIII lines observed in the solar polar coronal
  holes}}.
\bjtitle{\aap}
\bvolume{339},
\bfpage{208}.
\adsurl{1998A\%26A...339..208B}.
\end{barticle}
\endbibitem

\bibitem[\protect\citeauthoryear{{Benz} and
  {Wentzel}}{1981}]{1981A&A....94..100B}
\begin{barticle}
\bauthor{\bsnm{{Benz}}, \binits{A.O.}},
\bauthor{\bsnm{{Wentzel}}, \binits{D.G.}}:
\byear{1981},
\batitle{{Coronal evolution and solar type I radio bursts - an ion-acoustic
  wave model}}.
\bjtitle{\aap}
\bvolume{94},
\bfpage{100}.
\adsurl{1981A\%26A....94..100B}.
\end{barticle}
\endbibitem

\bibitem[\protect\citeauthoryear{{Bian} and
  {Kontar}}{2010}]{2010PhPl...17f2308B}
\begin{barticle}
\bauthor{\bsnm{{Bian}}, \binits{N.H.}},
\bauthor{\bsnm{{Kontar}}, \binits{E.P.}}:
\byear{2010},
\batitle{{A gyrofluid description of Alfv{\'e}nic turbulence and its parallel
  electric field}}.
\bjtitle{Physics of Plasmas}
\bvolume{17}(\bissue{6}),
\bfpage{062308}.
\doiurl{10.1063/1.3439682}.
\adsurl{2010PhPl...17f2308B}.
\end{barticle}
\endbibitem

\bibitem[\protect\citeauthoryear{{Bian} and
  {Kontar}}{2011}]{2011A&A...527A.130B}
\begin{barticle}
\bauthor{\bsnm{{Bian}}, \binits{N.H.}},
\bauthor{\bsnm{{Kontar}}, \binits{E.P.}}:
\byear{2011},
\batitle{{Parallel electric field amplification by phase mixing of Alfven
  waves}}.
\bjtitle{\aap}
\bvolume{527},
\bfpage{A130}.
\doiurl{10.1051/0004-6361/201015385}.
\adsurl{2011A\%26A...527A.130B}.
\end{barticle}
\endbibitem

\bibitem[\protect\citeauthoryear{{Bian} and
  {Tsiklauri}}{2009}]{2009PhPl...16f4503B}
\begin{barticle}
\bauthor{\bsnm{{Bian}}, \binits{N.H.}},
\bauthor{\bsnm{{Tsiklauri}}, \binits{D.}}:
\byear{2009},
\batitle{{Compressible Hall magnetohydrodynamics in a strong magnetic field}}.
\bjtitle{Physics of Plasmas}
\bvolume{16}(\bissue{6}),
\bfpage{064503}.
\doiurl{10.1063/1.3159862}.
\adsurl{2009PhPl...16f4503B}.
\end{barticle}
\endbibitem

\bibitem[\protect\citeauthoryear{{Bian}, {Kontar}, and
  {Brown}}{2010}]{2010A&A...519A.114B}
\begin{barticle}
\bauthor{\bsnm{{Bian}}, \binits{N.H.}},
\bauthor{\bsnm{{Kontar}}, \binits{E.P.}},
\bauthor{\bsnm{{Brown}}, \binits{J.C.}}:
\byear{2010},
\batitle{{Parallel electric field generation by Alfv{\'e}n wave turbulence}}.
\bjtitle{\aap}
\bvolume{519},
\bfpage{A114}.
\doiurl{10.1051/0004-6361/201014048}.
\adsurl{2010A\%26A...519A.114B}.
\end{barticle}
\endbibitem

\bibitem[\protect\citeauthoryear{{Bruno} and
  {Carbone}}{2013}]{2013LRSP...10....2B}
\begin{barticle}
\bauthor{\bsnm{{Bruno}}, \binits{R.}},
\bauthor{\bsnm{{Carbone}}, \binits{V.}}:
\byear{2013},
\batitle{{The Solar Wind as a Turbulence Laboratory}}.
\bjtitle{Living Reviews in Solar Physics}
\bvolume{10},
\bfpage{2}.
\doiurl{10.12942/lrsp-2013-2}.
\adsurl{2013LRSP...10....2B}.
\end{barticle}
\endbibitem

\bibitem[\protect\citeauthoryear{{Chaston}
  \textit{et~al.}}{2008}]{2008PhRvL.100q5003C}
\begin{barticle}
\bauthor{\bsnm{{Chaston}}, \binits{C.C.}},
\bauthor{\bsnm{{Salem}}, \binits{C.}},
\bauthor{\bsnm{{Bonnell}}, \binits{J.W.}},
\bauthor{\bsnm{{Carlson}}, \binits{C.W.}},
\bauthor{\bsnm{{Ergun}}, \binits{R.E.}},
\bauthor{\bsnm{{Strangeway}}, \binits{R.J.}},
\bauthor{\bsnm{{McFadden}}, \binits{J.P.}}:
\byear{2008},
\batitle{{The Turbulent Alfv{\'e}nic Aurora}}.
\bjtitle{Physical Review Letters}
\bvolume{100}(\bissue{17}),
\bfpage{175003}.
\doiurl{10.1103/PhysRevLett.100.175003}.
\adsurl{2008PhRvL.100q5003C}.
\end{barticle}
\endbibitem

\bibitem[\protect\citeauthoryear{{Chen}
  \textit{et~al.}}{2012}]{2012PhRvL.109c5001C}
\begin{barticle}
\bauthor{\bsnm{{Chen}}, \binits{C.H.K.}},
\bauthor{\bsnm{{Salem}}, \binits{C.S.}},
\bauthor{\bsnm{{Bonnell}}, \binits{J.W.}},
\bauthor{\bsnm{{Mozer}}, \binits{F.S.}},
\bauthor{\bsnm{{Bale}}, \binits{S.D.}}:
\byear{2012},
\batitle{{Density Fluctuation Spectrum of Solar Wind Turbulence between Ion and
  Electron Scales}}.
\bjtitle{Physical Review Letters}
\bvolume{109}(\bissue{3}),
\bfpage{035001}.
\doiurl{10.1103/PhysRevLett.109.035001}.
\adsurl{2012PhRvL.109c5001C}.
\end{barticle}
\endbibitem

\bibitem[\protect\citeauthoryear{{Dulk} and
  {McLean}}{1978}]{1978SoPh...57..279D}
\begin{barticle}
\bauthor{\bsnm{{Dulk}}, \binits{G.A.}},
\bauthor{\bsnm{{McLean}}, \binits{D.J.}}:
\byear{1978},
\batitle{{Coronal magnetic fields}}.
\bjtitle{\solphys}
\bvolume{57},
\bfpage{279}.
\doiurl{10.1007/BF00160102}.
\adsurl{1978SoPh...57..279D}.
\end{barticle}
\endbibitem

\bibitem[\protect\citeauthoryear{{Elgar{\o}y}}{1977}]{1977sns..book.....E}
\begin{bbook}
\bauthor{\bsnm{{Elgar{\o}y}}, \binits{E.{\O}.}}:
\byear{1977},
\bbtitle{{Solar noise storms, Pergamon Press, Oxford}}.
\adsurl{1977sns..book.....E}.
\end{bbook}
\endbibitem

\bibitem[\protect\citeauthoryear{{Goertz} and
  {Boswell}}{1979}]{1979JGR....84.7239G}
\begin{barticle}
\bauthor{\bsnm{{Goertz}}, \binits{C.K.}},
\bauthor{\bsnm{{Boswell}}, \binits{R.W.}}:
\byear{1979},
\batitle{{Magnetosphere-ionosphere coupling}}.
\bjtitle{\jgr}
\bvolume{84},
\bfpage{7239}.
\doiurl{10.1029/JA084iA12p07239}.
\adsurl{1979JGR....84.7239G}.
\end{barticle}
\endbibitem

\bibitem[\protect\citeauthoryear{{Gopalswamy}
  \textit{et~al.}}{1986}]{1986BAICz..37..115G}
\begin{barticle}
\bauthor{\bsnm{{Gopalswamy}}, \binits{N.}},
\bauthor{\bsnm{{Thejappa}}, \binits{G.}},
\bauthor{\bsnm{{Sastry}}, \binits{C.V.}},
\bauthor{\bsnm{{Tlamicha}}, \binits{A.}}:
\byear{1986},
\batitle{{Estimation of coronal magnetic fields using Type-I emission}}.
\bjtitle{Bulletin of the Astronomical Institutes of Czechoslovakia}
\bvolume{37},
\bfpage{115}.
\adsurl{1986BAICz..37..115G}.
\end{barticle}
\endbibitem

\bibitem[\protect\citeauthoryear{{Hamilton} and
  {Petrosian}}{1987}]{1987ApJ...321..721H}
\begin{barticle}
\bauthor{\bsnm{{Hamilton}}, \binits{R.J.}},
\bauthor{\bsnm{{Petrosian}}, \binits{V.}}:
\byear{1987},
\batitle{{Generation of plasma waves by thick-target electron beams, and the
  expected radiation signature}}.
\bjtitle{\apj}
\bvolume{321},
\bfpage{721}.
\doiurl{10.1086/165665}.
\adsurl{1987ApJ...321..721H}.
\end{barticle}
\endbibitem

\bibitem[\protect\citeauthoryear{{Hasegawa} and
  {Chen}}{1975}]{1975PhRvL..35..370H}
\begin{barticle}
\bauthor{\bsnm{{Hasegawa}}, \binits{A.}},
\bauthor{\bsnm{{Chen}}, \binits{L.}}:
\byear{1975},
\batitle{{Kinetic process of plasma heating due to Alfv{\'e}n wave
  excitation}}.
\bjtitle{Physical Review Letters}
\bvolume{35},
\bfpage{370}.
\doiurl{10.1103/PhysRevLett.35.370}.
\adsurl{1975PhRvL..35..370H}.
\end{barticle}
\endbibitem

\bibitem[\protect\citeauthoryear{{He}
  \textit{et~al.}}{2012}]{2012ApJ...745L...8H}
\begin{barticle}
\bauthor{\bsnm{{He}}, \binits{J.}},
\bauthor{\bsnm{{Tu}}, \binits{C.}},
\bauthor{\bsnm{{Marsch}}, \binits{E.}},
\bauthor{\bsnm{{Yao}}, \binits{S.}}:
\byear{2012},
\batitle{{Do Oblique Alfv{\'e}n/Ion-cyclotron or Fast-mode/Whistler Waves
  Dominate the Dissipation of Solar Wind Turbulence near the Proton Inertial
  Length?}}
\bjtitle{\apjl}
\bvolume{745},
\bfpage{L8}.
\doiurl{10.1088/2041-8205/745/1/L8}.
\adsurl{2012ApJ...745L...8H}.
\end{barticle}
\endbibitem

\bibitem[\protect\citeauthoryear{{Kontar}, {Ratcliffe}, and
  {Bian}}{2012}]{2012A&A...539A..43K}
\begin{barticle}
\bauthor{\bsnm{{Kontar}}, \binits{E.P.}},
\bauthor{\bsnm{{Ratcliffe}}, \binits{H.}},
\bauthor{\bsnm{{Bian}}, \binits{N.H.}}:
\byear{2012},
\batitle{{Wave-particle interactions in non-uniform plasma and the
  interpretation of hard X-ray spectra in solar flares}}.
\bjtitle{\aap}
\bvolume{539},
\bfpage{A43}.
\doiurl{10.1051/0004-6361/201118216}.
\adsurl{2012A\%26A...539A..43K}.
\end{barticle}
\endbibitem

\bibitem[\protect\citeauthoryear{{Lysak} and
  {Lotko}}{1996}]{1996JGR...101.5085L}
\begin{barticle}
\bauthor{\bsnm{{Lysak}}, \binits{R.L.}},
\bauthor{\bsnm{{Lotko}}, \binits{W.}}:
\byear{1996},
\batitle{{On the kinetic dispersion relation for shear Alfv{\'e}n waves}}.
\bjtitle{\jgr}
\bvolume{101},
\bfpage{5085}.
\doiurl{10.1029/95JA03712}.
\adsurl{1996JGR...101.5085L}.
\end{barticle}
\endbibitem

\bibitem[\protect\citeauthoryear{{Marsch} and {Tu}}{1990}]{1990JGR....9511945M}
\begin{barticle}
\bauthor{\bsnm{{Marsch}}, \binits{E.}},
\bauthor{\bsnm{{Tu}}, \binits{C.-Y.}}:
\byear{1990},
\batitle{{Spectral and spatial evolution of compressible turbulence in the
  inner solar wind}}.
\bjtitle{\jgr}
\bvolume{95},
\bfpage{11945}.
\doiurl{10.1029/JA095iA08p11945}.
\adsurl{1990JGR....9511945M}.
\end{barticle}
\endbibitem

\bibitem[\protect\citeauthoryear{{Matthaeus}
  \textit{et~al.}}{1999}]{1999ApJ...523L..93M}
\begin{barticle}
\bauthor{\bsnm{{Matthaeus}}, \binits{W.H.}},
\bauthor{\bsnm{{Zank}}, \binits{G.P.}},
\bauthor{\bsnm{{Oughton}}, \binits{S.}},
\bauthor{\bsnm{{Mullan}}, \binits{D.J.}},
\bauthor{\bsnm{{Dmitruk}}, \binits{P.}}:
\byear{1999},
\batitle{{Coronal Heating by Magnetohydrodynamic Turbulence Driven by Reflected
  Low-Frequency Waves}}.
\bjtitle{\apjl}
\bvolume{523},
\bfpage{L93}.
\doiurl{10.1086/312259}.
\adsurl{1999ApJ...523L..93M}.
\end{barticle}
\endbibitem

\bibitem[\protect\citeauthoryear{{McLean} and
  {Labrum}}{1985}]{1985srph.book.....M}
\begin{bbook}
\bauthor{\bsnm{{McLean}}, \binits{D.J.}},
\bauthor{\bsnm{{Labrum}}, \binits{N.R.}}:
\byear{1985},
\bbtitle{{Solar radiophysics: Studies of emission from the sun at metre
  wavelengths, Cambridge University Press}}.
\adsurl{1985srph.book.....M}.
\end{bbook}
\endbibitem

\bibitem[\protect\citeauthoryear{{Melrose}}{1980a}]{1980SoPh...67..357M}
\begin{barticle}
\bauthor{\bsnm{{Melrose}}, \binits{D.B.}}:
\byear{1980}a,
\batitle{{A plasma-emission mechanism for type I solar radio emission}}.
\bjtitle{\solphys}
\bvolume{67},
\bfpage{357}.
\doiurl{10.1007/BF00149813}.
\adsurl{1980SoPh...67..357M}.
\end{barticle}
\endbibitem

\bibitem[\protect\citeauthoryear{{Melrose}}{1980b}]{1980gbs..bookR....M}
\begin{bbook}
\bauthor{\bsnm{{Melrose}}, \binits{D.B.}}:
\byear{1980}b,
\bbtitle{{Plasma astrophysics: Nonthermal processes in diffuse magnetized
  plasmas. Volume 2 - Astrophysical applications, Gordon and Breach Science
  Publishers, New York}}.
\adsurl{1980gbs..bookR....M}.
\end{bbook}
\endbibitem

\bibitem[\protect\citeauthoryear{{Melrose} and
  {Stenhouse}}{1977}]{1977AuJPh..30..481M}
\begin{barticle}
\bauthor{\bsnm{{Melrose}}, \binits{D.B.}},
\bauthor{\bsnm{{Stenhouse}}, \binits{J.E.}}:
\byear{1977},
\batitle{{Emission and absorption of Langmuir waves by anisotropic unmagnetized
  particles}}.
\bjtitle{Australian Journal of Physics}
\bvolume{30},
\bfpage{481}.
\adsurl{1977AuJPh..30..481M}.
\end{barticle}
\endbibitem

\bibitem[\protect\citeauthoryear{{Melrose} and
  {Wheatland}}{2014}]{2014SoPh..289..881M}
\begin{barticle}
\bauthor{\bsnm{{Melrose}}, \binits{D.B.}},
\bauthor{\bsnm{{Wheatland}}, \binits{M.S.}}:
\byear{2014},
\batitle{{Bulk Energization of Electrons in Solar Flares by Alfv{\'e}n Waves}}.
\bjtitle{\solphys}
\bvolume{289},
\bfpage{881}.
\doiurl{10.1007/s11207-013-0376-7}.
\adsurl{2014SoPh..289..881M}.
\end{barticle}
\endbibitem

\bibitem[\protect\citeauthoryear{{Morton}, {Tomczyk}, and
  {Pinto}}{2015}]{2015NatCo...6E7813M}
\begin{barticle}
\bauthor{\bsnm{{Morton}}, \binits{R.J.}},
\bauthor{\bsnm{{Tomczyk}}, \binits{S.}},
\bauthor{\bsnm{{Pinto}}, \binits{R.}}:
\byear{2015},
\batitle{{Investigating Alfv{\'e}nic wave propagation in coronal open-field
  regions}}.
\bjtitle{Nature Communications}
\bvolume{6},
\bfpage{7813}.
\doiurl{10.1038/ncomms8813}.
\adsurl{2015NatCo...6E7813M}.
\end{barticle}
\endbibitem

\bibitem[\protect\citeauthoryear{{Pierrard} and
  {Voitenko}}{2013}]{2013SoPh..288..355P}
\begin{barticle}
\bauthor{\bsnm{{Pierrard}}, \binits{V.}},
\bauthor{\bsnm{{Voitenko}}, \binits{Y.}}:
\byear{2013},
\batitle{{Modification of Proton Velocity Distributions by Alfv{\'e}nic
  Turbulence in the Solar Wind}}.
\bjtitle{\solphys}
\bvolume{288},
\bfpage{355}.
\doiurl{10.1007/s11207-013-0294-8}.
\adsurl{2013SoPh..288..355P}.
\end{barticle}
\endbibitem

\bibitem[\protect\citeauthoryear{{Ratcliffe}}{2013}]{2013PhDT.......293R}
\begin{botherref}
\oauthor{\bsnm{{Ratcliffe}}, \binits{H.}}:
2013,
{Electron beam evolution and radio emission in the inhomogeneous solar corona}.
PhD thesis,
University of Glasgow (United Kingdom).
\adsurl{2013PhDT.......293R}.
\end{botherref}
\endbibitem

\bibitem[\protect\citeauthoryear{{Ratcliffe} and
  {Kontar}}{2014}]{2014A&A...562A..57R}
\begin{barticle}
\bauthor{\bsnm{{Ratcliffe}}, \binits{H.}},
\bauthor{\bsnm{{Kontar}}, \binits{E.P.}}:
\byear{2014},
\batitle{{Plasma radio emission from inhomogeneous collisional plasma of a
  flaring loop}}.
\bjtitle{\aap}
\bvolume{562},
\bfpage{A57}.
\doiurl{10.1051/0004-6361/201322263}.
\adsurl{2014A\%26A...562A..57R}.
\end{barticle}
\endbibitem

\bibitem[\protect\citeauthoryear{{Reid} and
  {Kontar}}{2010}]{2010ApJ...721..864R}
\begin{barticle}
\bauthor{\bsnm{{Reid}}, \binits{H.A.S.}},
\bauthor{\bsnm{{Kontar}}, \binits{E.P.}}:
\byear{2010},
\batitle{{Solar Wind Density Turbulence and Solar Flare Electron Transport from
  the Sun to the Earth}}.
\bjtitle{\apj}
\bvolume{721},
\bfpage{864}.
\doiurl{10.1088/0004-637X/721/1/864}.
\adsurl{2010ApJ...721..864R}.
\end{barticle}
\endbibitem

\bibitem[\protect\citeauthoryear{{Rudakov}
  \textit{et~al.}}{2012}]{2012PhPl...19d2704R}
\begin{barticle}
\bauthor{\bsnm{{Rudakov}}, \binits{L.}},
\bauthor{\bsnm{{Crabtree}}, \binits{C.}},
\bauthor{\bsnm{{Ganguli}}, \binits{G.}},
\bauthor{\bsnm{{Mithaiwala}}, \binits{M.}}:
\byear{2012},
\batitle{{Quasilinear evolution of plasma distribution functions and
  consequences on wave spectrum and perpendicular ion heating in the turbulent
  solar wind}}.
\bjtitle{Physics of Plasmas}
\bvolume{19}(\bissue{4}),
\bfpage{042704}.
\doiurl{10.1063/1.3698407}.
\adsurl{2012PhPl...19d2704R}.
\end{barticle}
\endbibitem

\bibitem[\protect\citeauthoryear{{Ryutov}}{1969}]{1969JETP...30..131R}
\begin{barticle}
\bauthor{\bsnm{{Ryutov}}, \binits{D.D.}}:
\byear{1969},
\batitle{{Quasilinear Relaxation of an Electron Beam in an Inhomogeneous
  Plasma}}.
\bjtitle{Soviet Journal of Experimental and Theoretical Physics}
\bvolume{30},
\bfpage{131}.
\adsurl{1969JETP...30..131R}.
\end{barticle}
\endbibitem

\bibitem[\protect\citeauthoryear{{Schad}
  \textit{et~al.}}{2016}]{2016ApJ...833....5S}
\begin{barticle}
\bauthor{\bsnm{{Schad}}, \binits{T.A.}},
\bauthor{\bsnm{{Penn}}, \binits{M.J.}},
\bauthor{\bsnm{{Lin}}, \binits{H.}},
\bauthor{\bsnm{{Judge}}, \binits{P.G.}}:
\byear{2016},
\batitle{{Vector Magnetic Field Measurements along a Cooled Stereo-imaged
  Coronal Loop}}.
\bjtitle{\apj}
\bvolume{833},
\bfpage{5}.
\doiurl{10.3847/0004-637X/833/1/5}.
\adsurl{2016ApJ...833....5S}.
\end{barticle}
\endbibitem

\bibitem[\protect\citeauthoryear{{Spicer}, {Benz}, and
  {Huba}}{1982}]{1982A&A...105..221S}
\begin{barticle}
\bauthor{\bsnm{{Spicer}}, \binits{D.S.}},
\bauthor{\bsnm{{Benz}}, \binits{A.O.}},
\bauthor{\bsnm{{Huba}}, \binits{J.D.}}:
\byear{1982},
\batitle{{Solar type I noise storms and newly emerging magnetic flux}}.
\bjtitle{\aap}
\bvolume{105},
\bfpage{221}.
\adsurl{1982A\%26A...105..221S}.
\end{barticle}
\endbibitem

\bibitem[\protect\citeauthoryear{{Stasiewicz}
  \textit{et~al.}}{2000}]{2000SSRv...92..423S}
\begin{barticle}
\bauthor{\bsnm{{Stasiewicz}}, \binits{K.}},
\bauthor{\bsnm{{Bellan}}, \binits{P.}},
\bauthor{\bsnm{{Chaston}}, \binits{C.}},
\bauthor{\bsnm{{Kletzing}}, \binits{C.}},
\bauthor{\bsnm{{Lysak}}, \binits{R.}},
\bauthor{\bsnm{{Maggs}}, \binits{J.}},
\bauthor{\bsnm{{Pokhotelov}}, \binits{O.}},
\bauthor{\bsnm{{Seyler}}, \binits{C.}},
\bauthor{\bsnm{{Shukla}}, \binits{P.}},
\bauthor{\bsnm{{Stenflo}}, \binits{L.}},
\bauthor{\bsnm{{Streltsov}}, \binits{A.}},
\bauthor{\bsnm{{Wahlund}}, \binits{J.-E.}}:
\byear{2000},
\batitle{{Small Scale Alfv{\'e}nic Structure in the Aurora}}.
\bjtitle{\ssr}
\bvolume{92},
\bfpage{423}.
\adsurl{2000SSRv...92..423S}.
\end{barticle}
\endbibitem

\bibitem[\protect\citeauthoryear{{Thejappa}}{1991}]{1991SoPh..132..173T}
\begin{barticle}
\bauthor{\bsnm{{Thejappa}}, \binits{G.}}:
\byear{1991},
\batitle{{A self-consistent model for the storm radio emission from the sun}}.
\bjtitle{\solphys}
\bvolume{132},
\bfpage{173}.
\doiurl{10.1007/BF00159137}.
\adsurl{1991SoPh..132..173T}.
\end{barticle}
\endbibitem

\bibitem[\protect\citeauthoryear{{Tsytovich} and {ter
  Haar}}{1995}]{1995lnlp.book.....T}
\begin{bbook}
\bauthor{\bsnm{{Tsytovich}}, \binits{V.N.}},
\bauthor{\bsnm{{ter Haar}}, \binits{D.}}:
\byear{1995},
\bbtitle{{Lectures on Non-linear Plasma Kinetics, Springer, Berlin}}.
\adsurl{1995lnlp.book.....T}.
\end{bbook}
\endbibitem

\bibitem[\protect\citeauthoryear{{Vedenov} and
  {Velikhov}}{1963}]{1963JETP...16..682V}
\begin{barticle}
\bauthor{\bsnm{{Vedenov}}, \binits{A.A.}},
\bauthor{\bsnm{{Velikhov}}, \binits{E.P.}}:
\byear{1963},
\batitle{{Quasilinear Approximation in the Kinetic Theory of a Low-density
  Plasma}}.
\bjtitle{Soviet Journal of Experimental and Theoretical Physics}
\bvolume{16},
\bfpage{682}.
\adsurl{1963JETP...16..682V}.
\end{barticle}
\endbibitem

\bibitem[\protect\citeauthoryear{{Voitenko} and {De
  Keyser}}{2016}]{2016ApJ...832L..20V}
\begin{barticle}
\bauthor{\bsnm{{Voitenko}}, \binits{Y.}},
\bauthor{\bsnm{{De Keyser}}, \binits{J.}}:
\byear{2016},
\batitle{{MHD-Kinetic Transition in Imbalanced Alfv{\'e}nic Turbulence}}.
\bjtitle{\apjl}
\bvolume{832},
\bfpage{L20}.
\doiurl{10.3847/2041-8205/832/2/L20}.
\adsurl{2016ApJ...832L..20V}.
\end{barticle}
\endbibitem

\bibitem[\protect\citeauthoryear{{Voitenko} and
  {Goossens}}{2000}]{2000A&A...357.1073V}
\begin{barticle}
\bauthor{\bsnm{{Voitenko}}, \binits{Y.}},
\bauthor{\bsnm{{Goossens}}, \binits{M.}}:
\byear{2000},
\batitle{{Nonlinear decay of phase-mixed Alfv{\'e}n waves in the solar
  corona}}.
\bjtitle{\aap}
\bvolume{357},
\bfpage{1073}.
\adsurl{2000A\%26A...357.1073V}.
\end{barticle}
\endbibitem

\bibitem[\protect\citeauthoryear{{Voitenko} and
  {Goossens}}{2006}]{2006SSRv..122..255V}
\begin{barticle}
\bauthor{\bsnm{{Voitenko}}, \binits{Y.}},
\bauthor{\bsnm{{Goossens}}, \binits{M.}}:
\byear{2006},
\batitle{{Energization of Plasma Species by Intermittent Kinetic Alfv{\'e}n
  Waves}}.
\bjtitle{\ssr}
\bvolume{122},
\bfpage{255}.
\doiurl{10.1007/s11214-006-8212-0}.
\adsurl{2006SSRv..122..255V}.
\end{barticle}
\endbibitem

\bibitem[\protect\citeauthoryear{{Voitenko} and
  {Pierrard}}{2013}]{2013SoPh..288..369V}
\begin{barticle}
\bauthor{\bsnm{{Voitenko}}, \binits{Y.}},
\bauthor{\bsnm{{Pierrard}}, \binits{V.}}:
\byear{2013},
\batitle{{Velocity-Space Proton Diffusion in the Solar Wind Turbulence}}.
\bjtitle{\solphys}
\bvolume{288},
\bfpage{369}.
\doiurl{10.1007/s11207-013-0296-6}.
\adsurl{2013SoPh..288..369V}.
\end{barticle}
\endbibitem

\bibitem[\protect\citeauthoryear{{Voitenko}}{1998}]{1998SoPh..182..411V}
\begin{barticle}
\bauthor{\bsnm{{Voitenko}}, \binits{Y.M.}}:
\byear{1998},
\batitle{{Excitation of Kinetic Alfv{\'e}n Waves in a Flaring Loop}}.
\bjtitle{\solphys}
\bvolume{182},
\bfpage{411}.
\doiurl{10.1023/A:1005049006572}.
\adsurl{1998SoPh..182..411V}.
\end{barticle}
\endbibitem

\bibitem[\protect\citeauthoryear{{Warmuth} and
  {Mann}}{2005}]{2005A&A...435.1123W}
\begin{barticle}
\bauthor{\bsnm{{Warmuth}}, \binits{A.}},
\bauthor{\bsnm{{Mann}}, \binits{G.}}:
\byear{2005},
\batitle{{A model of the Alfv{\'e}n speed in the solar corona}}.
\bjtitle{\aap}
\bvolume{435},
\bfpage{1123}.
\doiurl{10.1051/0004-6361:20042169}.
\adsurl{2005A\%26A...435.1123W}.
\end{barticle}
\endbibitem

\bibitem[\protect\citeauthoryear{{Wentzel}}{1986}]{1986SoPh..103..141W}
\begin{barticle}
\bauthor{\bsnm{{Wentzel}}, \binits{D.G.}}:
\byear{1986},
\batitle{{A theory for the solar type-I radio continuum}}.
\bjtitle{\solphys}
\bvolume{103},
\bfpage{141}.
\doiurl{10.1007/BF00154864}.
\adsurl{1986SoPh..103..141W}.
\end{barticle}
\endbibitem

\end{thebibliography}

\end{article}
\end{document}